\pdfoutput=1

\documentclass[10pt,journal,compsoc]{IEEEtran}

\ifCLASSOPTIONcompsoc
  \usepackage[nocompress]{cite}
\else
  \usepackage{cite}
\fi

\usepackage{cite}
\usepackage{amsmath,amssymb,amsfonts}
\usepackage{algorithmic}
\usepackage{graphicx}
\usepackage{textcomp}
\usepackage{xcolor}

\usepackage{times}

\usepackage{amssymb}
\usepackage{amssymb,amsmath}
\usepackage{psfrag}
\usepackage{epsfig}
\usepackage{graphicx,subfigure}
\usepackage{color}
\usepackage{bm}
\usepackage{epstopdf}
\usepackage{verbatim}
\usepackage{graphicx}   

\usepackage{textcomp}

\usepackage[ruled]{algorithm2e}
\usepackage{amsmath,amssymb}
\usepackage{amsthm}
\usepackage{latexsym,verbatim}
\usepackage{psfrag,epsfig}
\usepackage{graphicx,subfigure}
\usepackage{color}
\usepackage{epstopdf}
\usepackage{listings}
\usepackage{tikz}
\usepackage{url}
\usepackage[ruled]{algorithm2e}

\usepackage{booktabs}

\usepackage{ifluatex,ifxetex}
\ifluatex
\usepackage{fontspec} 
\else\ifxetex
\usepackage{fontspec} 
\else 
\usepackage[T1]{fontenc}
\usepackage[utf8]{inputenc} 
\fi\fi

\newtheorem{theorem}{Theorem}

\newtheorem{insight}{Insight}

\begin{document}

\title{Impact of Temporary Fork on the Evolution of Mining Pools in Blockchain Networks:\\ An Evolutionary Game Analysis}

\author{Canhui Chen,
        Xu Chen,~\IEEEmembership{Senior Member,~IEEE,}
        Jiangshan Yu, ~\IEEEmembership{Member,~IEEE,}
        Weigang Wu,~\IEEEmembership{Member,~IEEE, }
         and Di Wu,~\IEEEmembership{Senior Member,~IEEE} 
\IEEEcompsocitemizethanks{\IEEEcompsocthanksitem Canhui Chen, Xu Chen, Weigang Wu, and Di Wu are with School of Computer Science and Engineering, Sun Yat-sen University, Guangzhou, China.
\protect\\
E-mails: chench296@mail2.sysu.edu.cn, chenxu35@mail.sysu.edu.cn, wuweig@mail.sysu.edu.cn, wudi27@mail.sysu.edu.cn
\IEEEcompsocthanksitem Jiangshan Yu is with Faculty of Information Technology, Monash University, Melbourne, Australia.
\protect\\
E-mail: Jiangshan.Yu@monash.edu}
\thanks{This manuscript has been accepted by IEEE Transactions on Network Science and Engineering. Xu Chen is the corresponding author.}
}


\IEEEtitleabstractindextext{%
\begin{abstract}
	Temporary fork is a fundamental phenomenon in many blockchains with proof of work, and the analysis of temporary fork has recently drawn great attention. Different from existing efforts that focus on the blockchain system factors such as block size, network propagation delay or block generation speed, in this paper we explore a new key dimension of computing power from the miners' perspective.  Specifically, we first propose a detailed mathematical model to characterize the impact of computing power competition of the mining pools on the temporary fork. We also derive closed-form formula of the probability of temporary fork and the expected mining reward of a mining pool. To reveal the long-term trends on the computing power distributions over the competing mining pools, we then develop an evolutionary game framework based on the temporary fork modeling and accordingly characterize the set of stable evolution equilibriums. Both extensive numerical simulations and realistic blockchain data based evaluation provide evidence to support our theoretical models and discoveries.
\end{abstract}

}

\maketitle

\IEEEdisplaynontitleabstractindextext

\IEEEpeerreviewmaketitle

\IEEEraisesectionheading{\section{Introduction}\label{sec:introduction}}

\IEEEPARstart{S}{ince} the first blockchain application Bitcoin \cite{bitcoin} was launched, blockchain has attracted more and more attention. Blockchain has numerous benefits such as decentralization, persistence, pseudonymity, and auditability. Currently, blockchain has many application scenarios such as finance, supply chain and IoT services \cite{melanie2019blockchain,bhattacharya2019bindaas,chen2018stochastic,baza2019b,yazdinejad2019blockchain}, many of which greatly depends on the decentralization feature of blockchain. 

The Proof-of-Work (PoW) is the most widely adopted consensus algorithm in blockchain networks such as Bitcoin and Ethereum. By solving math puzzles usually involving hash calculations, those who find a solution that meets the conditions are allowed to generate new blocks and earn a block reward. This economic incentive encourages participants (a.k.a. miners) to contribute their computing power as much as possible in solving PoW puzzles, which is often called mining. As the blockchain mining has become more complex and difficult, mining pools have been created to consolidate resources to form a single entity that mines blockchain blocks with collective computational power of all the miners in the pool, which leads to the formation of several large mining pools and the increase of the centralization in blockchain systems. The centralization crisis of the PoW-based blcokchain severely threatens these blockchain-based applications.

Investigating the long-term trends on decentralization remains a challenging issue \cite{gencer2018decentralization}. Motivated by this, we are going to study the evolution of mining pools in blockchain networks and reveal the long-term trends on the computing power distributions over the competing mining pools. We also investigate how the system settings, such as the network propagation delay, the block size, and the ratio of uncle block reward, affect the degree of decentralization of blockchain networks through the temporary fork perspective.

Temporary fork is a common fundamental phenomenon in many blockchain systems, and it happens when the miners discover a block at the same time, which would result in  split competing chains. And temporary forks are typically resolved in PoW blockchains such as Bitcoin when miners select which chain to form subsequent blocks upon \cite{asoloblockchain}.  In general, a higher probability of temporary fork implies that inconsistency occurs more frequently in the consensus, leading to a higher ratio of abandoned and stale blocks in the blockchain, which can be very harmful for the system performance such as transaction throughput and resource efficiency \cite{neudecker2019short}.  Motivated by this, the analysis of temporary fork has been drawing great research attention \cite{xiao2020survey}. Nevertheless, existing research efforts mainly focus on the impacting factors from the blockchain system point of view, such as block size, network propagation delay or block generation speed \cite{bitcoin_backbone, network_propagation}.

Along a different line, in this paper we explore a new key dimension of computing power for mining from the miners' perspective, which is much less understood yet and is critical for revealing the long-term mining patterns and evalution states of blockchains \cite{xiao2020survey}. Specifically, we first propose a detailed model of temporary fork in which the probability of stale blocks mined by a mining pool depends on its own hash rate and the hash rates of the other pools in the entire blockchain network. We show that the expected block reward received by a mining pool is not proportional to its computational power. Particularly, the pools with a large hash rate will have a smaller probability of mining a stale block, thus they are at an advantage in competition, and can get more block rewards, while a pool with small hash rate will be trapped into a disadvantaged state in competition. 
Therefore, miners may join a larger mining pool for higher benefits, thereby reducing the degree of decentralization of the blockchain network. Additionally, the uncle block reward is considered in our model, which was once utilized in the Ethereum \cite{ether}. Combining the mining reward analysis, we show that increasing the uncle block reward can help to mitigate the negative impact of temporary fork.

Furthermore, based on the proposed temporary fork model, we investigate the competition evolution dynamics of the mining pools in the blockchain as an evolutionary game. We theoretically characterize the stable equilibriums of the evolutionary game under various conditions, which are useful to reveal the long-term trends on the computing power distributions over the mining pools. We find that  when considering the temporary fork competition, increasing the uncle block reward helps to reduce the degree of centralization of computing power distributions in the blockchain network. Besides, we show that a larger network delay or a shorter block generation interval will lead to a higher probability of temporary fork, which will trap the mining pools with the small hash rates at a disadvantage in competition, leading the blockchain network to be more centralized in mining power distributions. Several numerical simulations are carried out to verify our theoretical analysis. Moreover, the distributions of mining power in four major realistic blockchain networks are compared and investigated, which are consistent with the theoretical results of our models.
	
	Our main contributions in this paper are summarized as follows.
	\begin{itemize}
	    \item We propose a detailed mathematical model to characterize the impact of computing power competition of the mining pools on the temporary fork in the blockchain network, and then derive closed-form formula of the probability of temporary fork and the expected mining reward of a mining pool.
	    
	    \item We present an evolutionary game framework based on the temporary fork modeling to analyze the evolution equilibrium of the mining pool competitions, and accordingly characterize the set of stable equilibriums.
	    
	    \item Based on the proposed models and the derived theoretical results, we obtain several key insights and reasonable explanations for the observed phenomenons in practical blockchain systems through both numerical and realistic data based evaluations.
	\end{itemize}

	For convenience, the main notations used in the paper are listed in Table \ref{notation_table}. 
	Besides, the rest of this paper is organized as follows. In Section 2, we put forward a detailed model of the temporary fork in the blockchain network and formulate the disproportionate mining reward of different mining pools. In Section 3, we develop an evolutionary game based on the aforementioned temporary fork model and investigate the evolution equilibrium. In Section 4, we review related literature. Section 5 concludes the paper with final remarks.
	
%

\section{Temporary Fork Model in Blockchain Network}

We consider a blockchain network in which there is a large population of $N$ miners distributed in $M$ mining pools, where $N\geq M$. Let $\bm{h}=\left[h_{1}, \ldots, h_{M}\right]^{\top}$ denote the hash rate vector and $V = \sum_{i=1}^{M}{h_{i}}$ denote the total hash rate, and let $\bm{x} =\left[x_{1}, \ldots, x_{M}\right]^{\top}$ denote the hash rate fractions (or normalized hash rates) of the pools, where $x_i = h_{i} / V$. 

When two valid but conflict blocks are generated at the same height of
the blockchain, only one of them will be accepted by the blockchain
and the other is called a stale block. We call the miner of the
accepted block a winner in this competition. 
For Bitcoin-like systems, the creator of the stale blocks will get no reward. While in some other blockchain network like Ethereum, which adopts the GHOST rule \cite{ghost}, the stale blocks may be considered to be the uncle block and can get a partial reward. To investigate the impact of the temporary fork, we consider the general case and denote $R$ as the block reward and $\theta R$ denotes the reward for the uncle block where $\theta \in [0,1)$.
In particular, $\theta = 0$ means that the uncle block reward is zero in the blockchain network, which is similar to the stale block case in Bitcoin.

Based on the setting above, the temporary fork model is discussed in detail in the following subsections.

\subsection{Temporary Fork Modeling}\label{TFM}

Temporary forks are mainly caused by network delays. The block propagation time includes transmission delay and transaction verification time. For a block of size $s$, the transmission delay can be modeled as $\tau_p(s) = \frac{s}{\gamma c}$, where $\gamma$ is the network scale-related parameter and $c$ is the average effective  bandwidth of each network link \cite{network_propagation,rizun2015transaction}. And the block verification time can be modeled as a linear function $\tau_v(s) = \beta s$, where $\beta$ is a parameter determined by both the network scale (i.e., number of nodes in the blockchain network) and the average verification speed of each node. Then the propagation delay for a block of size $s$ is 
\begin{equation}\nonumber
\tau(s) = \tau_p(s) + \tau_v(s) = \frac{s}{\gamma c} + \beta s.
\end{equation}
Note that the focus of our study is to model the impact of diversified computing powers of the mining pools on the temporary fork. Thus we model the propagation delay from the holistic network point of view, and hence assume a homogeneous network propagation delay across the mining pools to represent its impact.

\begin{table}[t]
	\caption{Summary of Notations}	\label{notation_table}
	\begin{tabular}{@{}ll@{}}
		\toprule
		Notation                & Description          \\ \midrule
		$x_i$                  & Hash rate fraction (or normalized hash rate) of mining pool $i$  \\
		$N$                    & Number of miners                  \\
		$M$                    & Number of mining pools         \\
		$R$                    & Block reward                         \\
		$\theta$               & Fraction of the uncle block reward         \\
		$s$                    & Block size                           \\
		$\tau(s)$              & Propagation delay for a block of size $s$    \\
		$\lambda$              & Block production rate            \\
		$P^{\text{fork}}_{i}$  & Probability of forking for mining pool $i$            \\ 
		$P^{\text{fail}}_{i}$  & Probability of failure in competition for mining pool $i$ \\
		$P^{\text{uncle}}_{i}$ & Probability that mining pool $i$ mines a uncle block                                \\
		$Y_i$                  & Expected mining reward of mining pool $i$                                                    \\
		$\omega_i$             & Hash rate specification of mining pool $i$   \\
		$r_i$				   & Population fraction of mining pool $i$ \\
		$p$					   & Expenditure for mining energy \\
		$y_i$                  & Expected payoff of a miner in pool $i$ \\
		\midrule                                    
	\end{tabular}
\end{table}

Similar to previous studies such as \cite{bitcoin,bitcoin_backbone}, we assume that the event of generating a new block in the blockchain network follows the Poisson process and hence the time duration between two consecutive blocks follows the exponential distribution, with a mean time $T$, i.e., the average block generation rate $\lambda = 1 /T$. The probability density function (pdf) of the time duration $t$ between two consecutive blocks can be formulated as
\begin{equation}\nonumber
f(t ; \lambda)=\left\{\begin{array}{ll}{\lambda e^{-\lambda t},} & {t \geq 0}, \\ 
{0,} & {\text { otherwise. }}\end{array}\right. 
\end{equation}
And the corresponding cumulative distribution function (cdf) can be integrated from the exponential pdf and obtained as follows
\begin{equation}\nonumber
F(t ; \lambda)=\left\{\begin{array}{ll}{0,} & {t<0}, \\ {1-e^{-\lambda t},} & {t \geq 0.}\end{array}\right. 
\end{equation}
Then the probability of  generating two blocks during a block confirmation period due to network propagation delay, noted as $P^{\Delta}$, can be formulated as
\begin{equation}\nonumber
P^{\Delta} = F(\tau(s) ; \lambda) = 1 - e^{-\lambda \tau(s)}.
\end{equation}
It is worth mentioning that $P^{\Delta}$ is not equal to the probability of fork. In fact, after mining the block, the miner that mined the block will continue to mine the next block, and only other miners find another block within $\Delta t$ will result in a fork. This phenomenon is called ``Last Block Effect''\cite{more_uncle_statistics}, i.e., the miner that produced the last block ``find out'' the block immediately rather than after waiting for $\tau(s)$ seconds for it to propagate through the network, and thus gains an advantage in decreasing his fork rate and finding the next block. 

\begin{figure}[!t]
	\centering
	\subfigure[Last Block Effect]{
		\includegraphics[width=3.1in,height=1.0in]{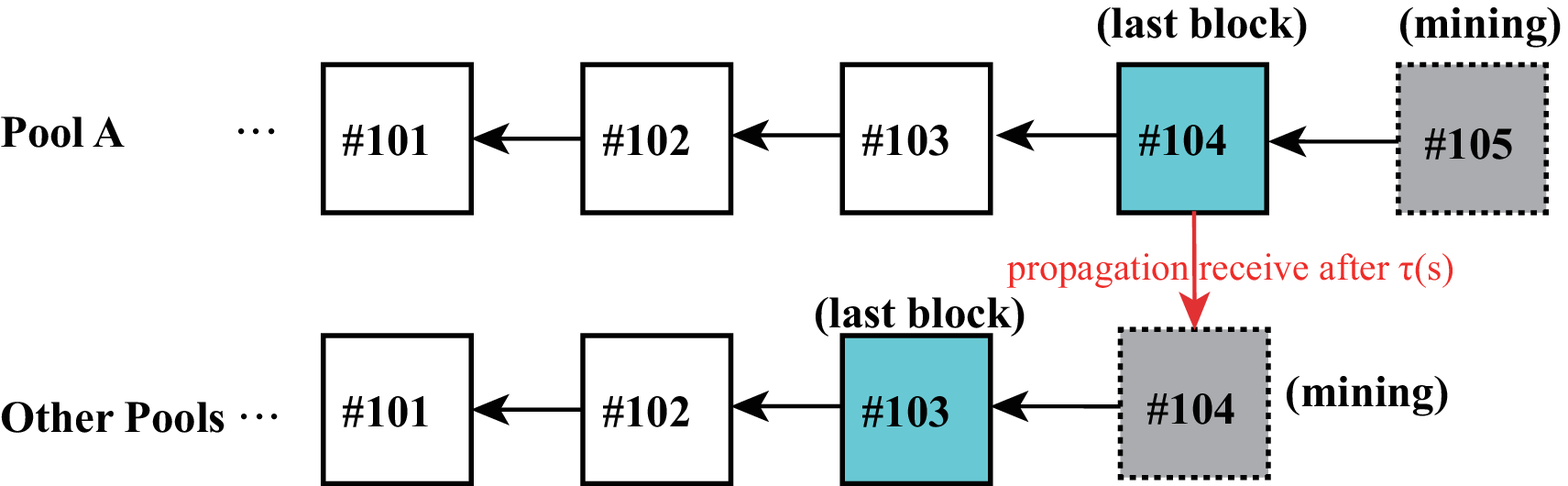}}
	\subfigure[Temporary fork and block inconsistencies]{
		\includegraphics[width=3.1in,height=1.2in]{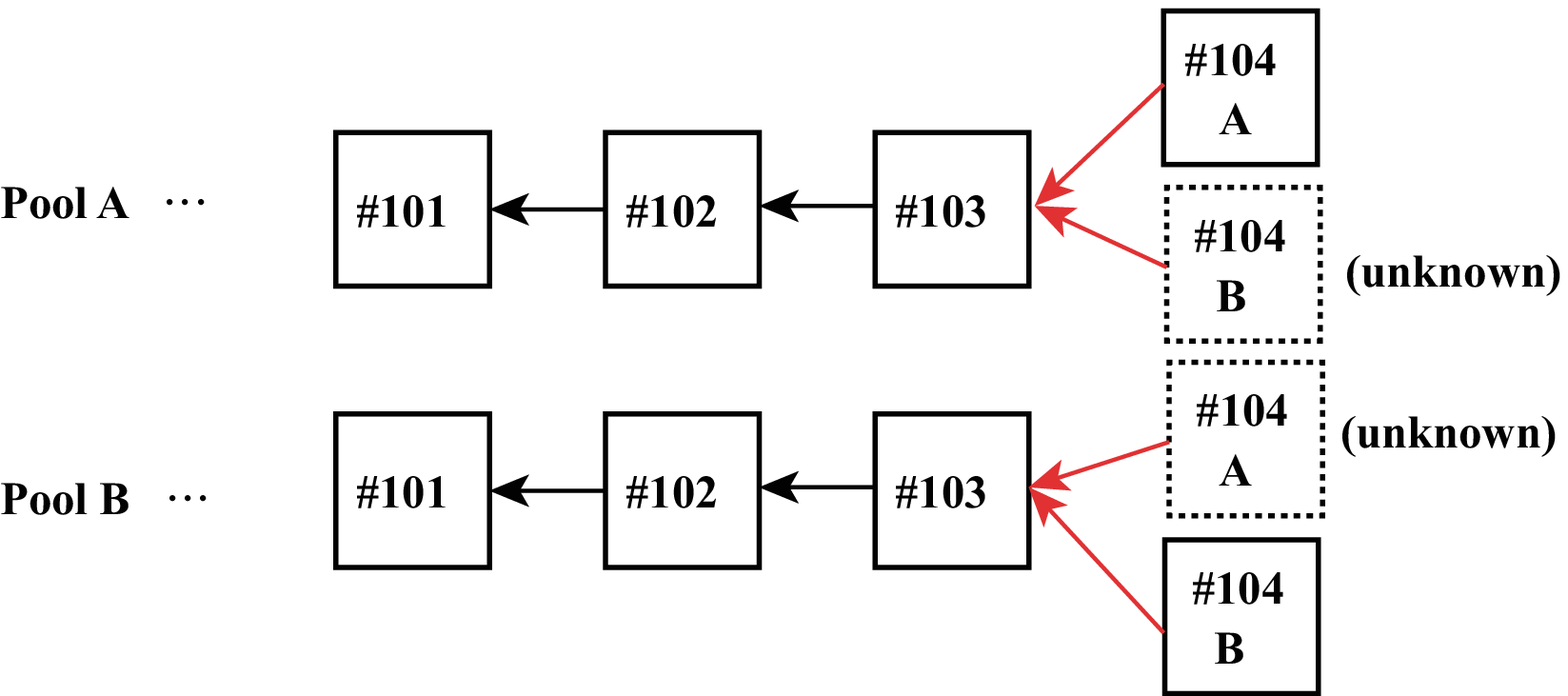}}
	\subfigure[Eventually, the longer chain will survive and one of the mining pools will win ]{\includegraphics[width=3.1in,height=1.4in]{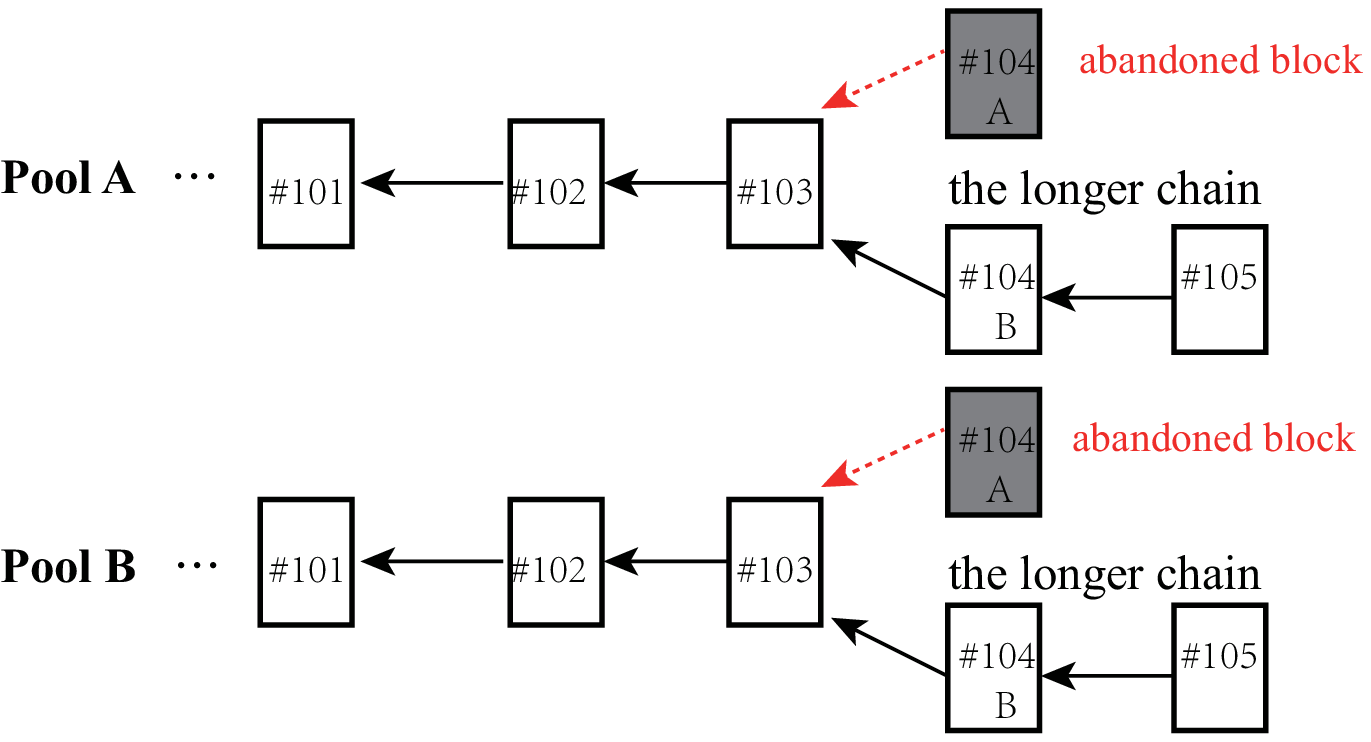}}
	\hfil
	\caption{Illustration of Temporary Fork}
	\label{temporary_fork}
\end{figure}

As shown in Figure \ref{temporary_fork} (a), mining Pool $A$ mines a new block \#104 at time $t$, which is the last block for Pool $A$, so Pool $A$ can continue to mine on the next block \#105 without waiting for $\tau(s)$ seconds for it to propagate through the network. However, the other pools in the blockchain network do not receive the new block \#104 before $t+ \tau(s)$, so they will continue to mine for the potential block \#104, as the block \#103 is the last block for them. If one of the other pools, i.e., Pool $B$, mines a block at the same height of 104 during the block propagation period, the temporary fork will happen due to the block inconsistencies, as shown in Figure \ref{temporary_fork} (b). Eventually, the longer chain will survive and one of the mining pool will win.

When a new block is generated into the blockchain network, since the network delay is the same for the miners, then the probability that a miner can successfully win the mining competition and find the block is proportional to its hash rate. Thus the probability of fork is related to the hash rate of the miner who finds the latest block. Therefore the probability that fork happens after a miner of mining pool $i$ finding the latest block is 
\begin{equation}\nonumber
P^{\text{fork}}_i = (1 - x_i) P^{\Delta} = (1 - x_i) (1 - e^{-\lambda \tau(s)}).
\end{equation}
And the probability that there is no fork after the miner in pool $i$ finding a block therefore is $P^{\text{chain}}_i = 1 - P^{\text{fork}}_i = (1 - x_i) e^{-\lambda \tau} \!-\! x_i$.

Generally, the faster the block generation speed, the higher the
frequency of temporary forks. In fact, the frequency of temporary fork
in Bitcoin is quite lower than that of Ethereum. Temporary forks will
result in two split competing branches. The algorithm of selecting of a
canonical chain is application specific. For example, Bitcoin selects
the chain contains most of the work (a.k.a. selecting the longest
chain) and GHOST selects the heaviest
subtree~\cite{natoli2019deconstructing}. 
In this paper, we only focus on the systems deploying the longest chain selection rule, as presented in Figure \ref{temporary_fork}.



For the blockchain following the longest chain rule, when a temporary fork appears, miners in the blockchain system need to choose one of the branches to mine until one of the chains becomes the longest chain. To account for this fact, we then propose a block selection algorithm shown in Algorithm \ref{minig_strategt} to model the chain selection behaviors. The algorithm is straightforward, when the temporary fork happens, the mining pools causing the temporary fork will continue to mine on their own chain, while the miners in those irrelevant mining pools will randomly select a branch to mine. This is motivated by the fact that many existing mining programs like the Satoshi client \cite{skudnov2012bitcoin} in bitcoin will simply pick the one it sees first to continue mining. Due to the dynamic fluctuations of the network conditions of the miners in reality, the first block of a fork that is propagated to a miner in irrelevant mining pools can be arbitrary and hence the chain selection of a miner in irrelevant mining pools is modeled as a random choice. Note that similar to existing studies \cite{bitcoin_backbone, network_propagation}, in this modeling study we do not consider the selfish mining attacks by some minority miners \cite{kkedziora2019analysis} to enable tractable analysis.


\begin{algorithm}[!t]
	\caption{Block selection algorithm}\label{minig_strategt}
	\KwIn{public chain for a mining pool} 	
	\While{Nobody finds a new block}{
		continue to mine on chain $c_1$
	}
	\If{My Pool finds a block on chain $c_1$}{
		publish the block immediately \\
		listen to the network while mining. \\
		\If{Another pool publishes the block during a block confirmation period leading to fork a new branch $c_2$}{
			\While{$c_1$.length $=$ $c_2$.length}{
				continue to mine on chain $c_1$
			}
			\If{$c_1$.length $>$ $c_2$.length}{
				My pool wins due to the lead of 1 block \\
				Reward $\leftarrow$ Reward + $R$
			}
			\Else{
				My Pool loses the competition \\
				Reward $\leftarrow$ Reward + $\theta R$
			}
		}
	}
	\If{Others find a new block}{	
		Attach the block to the current chain\\
		\If{Temporary fork happens}{
			Randomly select a chain branch to mine
		}
	}
\end{algorithm}

\subsection{Mining Reward Analysis}


Based on the temporary fork model and the block selection algorithm above, we then conduct a probabilistic analysis on the expected reward for a mining pool $i$. For simplification, as illustrated in Figure \ref{temporary_fork}, we will focus on the most common case that the chain is forked into two competing branches and the longer branch will win the competition\footnote{Our statistics (which is provided in Appendix A) shows that more than 95\% of the temporary forks in Ethereum are with exactly two competing branches, and similar observation holds for Bitcoin \cite{neudecker2019short}.}. The analysis of the case with more than two simultaneous branches is mathematically much more involved, and we believe the case with two branches can be a good approximation to provide useful insights. Also, since the analysis of double spending attacks is not the focus of this study, to simplify the analysis we do not require a branch to have multiple leading blocks to win.


Suppose that a mining pool $i$ just finds a new block $A$, leading to a new chain $c_1$. As shown in Section \ref{TFM}, then the probability of another pool $\tilde{i}$ to find another new block $B$ during the block confirmation period of block $A$ is $P^{\text{fork}}_i = (1 - x_i)(1 - e^{-\lambda \tau(s)})$, leading to another competing chain $c_2$ that forms the temporary fork. 

Let $\alpha$ and $\beta$ denote the hash rate fractions of these two competing mining pools $i$ and $\tilde{i}$, respectively. 
According to the block selection algorithm mentioned above, the mining pool $i$ will continue to mine on the $c_1$ until $c_1.\text{length} \neq c_2.\text{length}$, and similarly for mining pool $\tilde{i}$. The other mining pools except pools $i$ and $\tilde{i}$ will randomly select the chain $c_k, k \in \{1, 2\}$ to mine. For simplicity, we define that 
\begin{equation}\nonumber
\delta_k(j) = \begin{cases}
1, & \text{mining pool } i \text{ mines on chain } c_k, \\
0, & \text{otherwise}.
\end{cases}
\end{equation}
Then we have
\begin{equation}\nonumber
\mathbb{E} (\delta_1(j)) = \begin{cases}
1, & j = i, \\
0, & j = \tilde{i}, \\
\frac{1}{2}, & \text{otherwise}.
\end{cases}
\quad
\mathbb{E} (\delta_2(j)) = \begin{cases}
1, & j = \tilde{i}, \\
0, & j = i, \\
\frac{1}{2}, & \text{otherwise}.
\end{cases}
\end{equation}
Therefore, the mathematical expectation of the total hash rate on chain $c_1$ is
\begin{equation}\nonumber
\begin{aligned}
\mathbb{E}(\sum_{j=1}^{M} x_j \delta_1(j) ) &= \mathbb{E}(x_i + \!\!\!\! \sum_{j \neq i, j \neq \tilde{i}} \!\!\!\! x_j \delta_1(j) ) = x_i + \!\!\!\! \sum_{j \neq i, j \neq \tilde{i}} \!\!\!\! x_j \mathbb{E}(\delta_1(j)) \\ 
&= x_i + \frac{1}{2}\!\!\!\! \sum_{j \neq i, j \neq \tilde{i}} \!\!\!\! x_j = \alpha + \frac{1}{2}(1 - \alpha - \beta) \\
&= \frac{1+\alpha - \beta}{2}.
\end{aligned}
\end{equation}
Similarly, the mathematical expectation of the total hash rate on chain $c_2$ is
\begin{equation}\nonumber
\begin{aligned}
\mathbb{E}(\sum_{j=1}^{M} x_j \delta_2(j) ) &= \mathbb{E}(x_{\tilde{i}} + \!\!\!\! \sum_{j \neq i, j \neq \tilde{i}} \!\!\!\! x_j \delta_2(j) ) = x_{\tilde{i}} + \!\!\!\! \sum_{j \neq i, j \neq \tilde{i}} \!\!\!\! x_j \mathbb{E}(\delta_2(j)) \\ 
&= x_{\tilde{i}} + \frac{1}{2}\!\!\!\! \sum_{j \neq i, j \neq \tilde{i}} \!\!\!\! x_j = \beta + \frac{1}{2}(1 - \alpha - \beta) \\
&= \frac{1 - \alpha + \beta}{2}.
\end{aligned}
\end{equation}
In this case, when the next new block is generated, the probability that it appears (i.e., is found by a miner) on a chain $c_k,k \in \{1, 2\}$ equals to the hash rate fraction of the chain. Since a block is generated according to the Poisson distribution with the speed $\lambda$, the time duration $t_k$ for the next block to appear on a chain $k \in \{1, 2\}$ thus still follows the exponential distribution, but with different parameters $\lambda_k, k \in \{1, 2\}$ as follows:
\begin{equation}\nonumber
f(t ; \lambda_k)=\left\{\begin{array}{ll}{\lambda_k e^{-\lambda_k t},} & {t \geq 0,} \\ {0,} & {\text { otherwise, }}\end{array}\right.
\end{equation}
where $\lambda_1 = \frac{(1 + \alpha - \beta) \lambda}{2}$ and $\lambda_2 = \frac{(1 - \alpha + \beta)\lambda}{2}$.


Let $z = t_2 - t_1$ be a random variable describing the time difference of the next block appearing time on chains $c_2$ and $c_1$. Therefore, the probability density function (pdf) of $z$ is 
\begin{equation}\nonumber
f_z(z)=\left\{\begin{array}{ll}{	\frac{\lambda_1 \lambda_2}{\lambda_1 + \lambda_2}e^{-\lambda_2 z},} & {z \geq 0,} \\ {\frac{\lambda_1 \lambda_2}{\lambda_1 + \lambda_2}e^{\lambda_1 z},} & { { z < 0. }}\end{array}\right.
\end{equation}
And the cumulative distribution function (cdf) of $z$ is obtained as 
\begin{equation}\nonumber
F_z(z)=\left\{\begin{array}{ll}{1 - \frac{\lambda_1}{\lambda_1 + \lambda_2} e^{-\lambda_2 z},} & {z \geq 0,} \\ {\frac{\lambda_2}{\lambda_1 + \lambda_2} e^{\lambda_1 z},} & { { z < 0. }}\end{array}\right.
\end{equation}
Accordingly, the competing chain $c_2$ will win during the temporary fork if the appearing time of the next block on chain $c_2$ is ahead of the appearing time of the next block on chain $c_1$ by at least a confirmation period duration $\tau(s)$ (i.e., $t_1\geq t_2+\tau(s)$), which results in the block $A$ on the branching chain $c_1$ becoming a stale block. And we can derive the corresponding probability as  
\begin{equation}\nonumber
P(c_2 \text{ wins in next period}) = F_z(-\tau(s)) = \eta_1 e^{-\lambda \tau(s)  \eta_2},
\end{equation}
where $\eta_1 = \frac{1 - \alpha + \beta}{2}$ and $\eta_2 = \frac{1 + \alpha - \beta}{2}$.
Similarly, the probability that $c_1$ wins, resulting in the block $B$ on chain $c_2$ to be the stale block is that 
\begin{equation}\nonumber
P(c_1 \text{ wins in next period}) = 1 - F_z(\tau(s)) = \eta_2 e^{-\lambda \tau(s) \eta_1},
\end{equation}
and the probability that $c_1$ and $c_2$ tie in the next period is 
\begin{equation}\nonumber
P(\text{tie}) = 1 - \eta_1 e^{-\lambda \tau(s) \eta_2} - \eta_2 e^{-\lambda \tau(s) \eta_1}.
\end{equation}

If $c_1$ and $c_2$ tie in the first period, due to the adjustment of mining power over these two chain branches according to the block selection algorithm, the probability that either $c_1$ or $c_2$ eventually wins the competition is approximately 50\%. Therefore, the probability that block $A$ of pool $i$ will be a stale block due to the competition by pool $\tilde{i}$ can be formulated as
\begin{equation}\nonumber
P_{i,\tilde{i}}^{\text{fail}} = \eta_1 e^{-\lambda \tau(s) \eta_2} + \frac{1}{2} P(\text{tie}) .
\end{equation}

\begin{figure}[!htp]
	\centering
	\includegraphics[width=2.4in,height=2.0in]{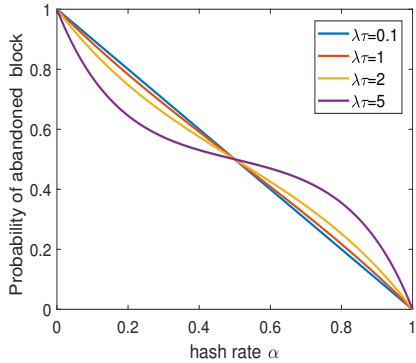}
	\caption{Probability Function}\label{uncle_rate_function}
\end{figure}

Since during a temporary fork the competing pool $\tilde{i}$ with pool $i$ can be any other pool except $i$, we have the competing hash rate fraction of $\beta \in \{x_1, \cdots, x_{i-1}, x_{i+1}, \cdots x_n\}$, and the corresponding probability of $P(\beta = x_j) = \frac{x_j}{1 - x_i}$. In this case, the probability that a block of a pool $i$ will be stale during a temporary fork is
\begin{equation}\nonumber
P^{\text{fail}}_i = \sum_{j \neq i}{\frac{x_j}{2(1 - x_i)} (1 + \eta_1 e^{- \lambda \tau(s) \eta_2} - \eta_2 e^{-\lambda \tau(s) \eta_1})},
\end{equation}
where we have $\eta_1 = \frac{1 - x_i + x_j}{2}$ and $\eta_2 = \frac{1 + x_i - x_j}{2}$ accordingly.
For simplification, the formula above can be approximated by the following formula using Taylor expansion as
\begin{equation}\nonumber
P^{\text{fail}}_i \approx \sum_{j \neq i}{\frac{x_j (1 - x_i + x_j)}{2(1 - x_i)}}.
\end{equation}
As an illustration, here we put forward a simple model for two mining pools based on the above analysis to intuitively explain the rationality of the approximation. Suppose there are only two mining pools $i$ and $j$ with hash rate $\alpha$ and $\beta$, subject to $\alpha + \beta = 1$. Thus, the probability for mining pool $i$ to mine a stale is 
\begin{eqnarray}\nonumber
\begin{aligned}
P^{\text{fail}}_i &= 1/2 * (1 + (1 - \alpha) e^{-\lambda \tau(s) \alpha} - \alpha e^{-\lambda \tau(s) (1-\alpha))} \\
& \approx 1 - \alpha .
\end{aligned}
\end{eqnarray}
As can be seen from the the actual probability function in Figure \ref{uncle_rate_function}, this approximation is reasonable, since generally, we have $\tau(s) < T$ in the blockchain network and it implies $\lambda\tau(s)<1$, which is close to the linear form.

Based on the analysis above, the probability for mining pool $i$ to mine a stale block is thus as
\begin{equation}\label{p_uncle}
P^{\text{uncle}}_i = P^{\text{fork}}_i * P^{\text{fail}}_i = \frac{1 - e^{-\lambda \tau(s)}}{2} \sum_{j \neq i}{x_j (1 - x_i + x_j)}.
\end{equation}
And accordingly the expected reward of mining pool $i$ can be derived by the following theorem.

\begin{theorem}\label{pool_reward}
	For a blockchain system with $M$ mining pools, if the hash rate ratio of the mining pool $i$ is $x_i$ ($i \in \{1, \cdots, M\}$), the basic block reward is $R$, the block generation rate is $\lambda$, the network delay is $\tau(s)$, and the uncle block reward is $\theta R$, the expected mining reward of a mining pool $i$ can be expressed as
	\begin{equation}\nonumber
	Y_i = x_i R( 1 - (1 - \theta) P^{\text{uncle}}_i ) ,
	\end{equation}
	where $P^{\text{uncle}}_i$ is given in (\ref{p_uncle}).
\end{theorem}

The proof is given in Appendix B. From the theorem above, we can observe that with a faster speed $\lambda$ that a block is generated or a larger network delay $\tau(s)$ on the blockchain, we have a larger  probability of a temporary fork with more stale or uncle blocks being generated (i.e., $P^{\text{uncle}}_i$ becomes larger). Second, we see that the expected reward of a mining pool does not grow linearly with its computing power. Instead, with a higher computing power $x_i$, a mining poor $i$ can enjoy a more significant shrinking in $P^{\text{uncle}}_i$ (i.e., the probability of generating stale or uncle blocks reduces more significantly), leading to a more significant advantage in gaining rewards.
This implies that concentrating in the mining pool can get a lower percentage of stale blocks, and thus get higher returns. Last but not least, we see that increasing the uncle reward ratio $\theta$ can help to mitigate the impact of stale or uncle blocks on the mining reward.



\subsection{Practical Insights}

We now apply our theoretical results  to gain some useful insights through numerical study and realistic blockchain data evaluation.

\begin{figure}[!htp]
	\centering
	\includegraphics[width=2.4in,height=2.0in]{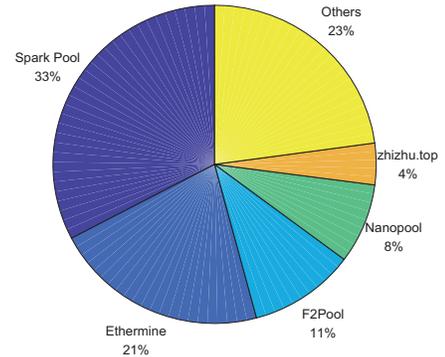}
	\caption{The top 5 mining pools' hash rates in Ethereum (2020.01)}\label{hash_rate_pie}
\end{figure}

\begin{figure*}[!htb]
	\centering
	\subfigure[Reward ratios for different network delay $\lambda \tau(s)$ with $\theta = 0$]{
		\includegraphics[width=2in,height=1.5in]{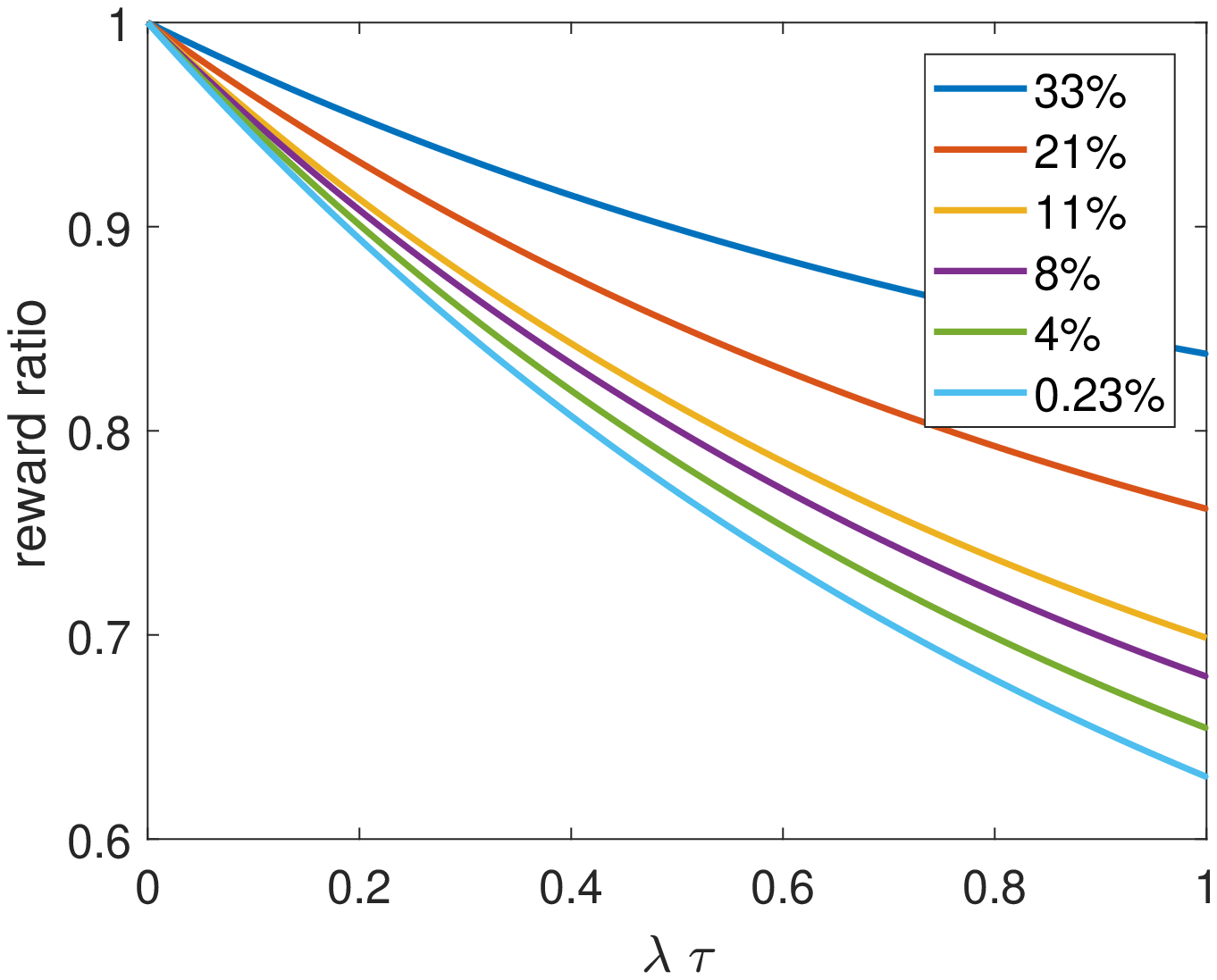}}
	\hfil
	\centering
	\subfigure[Reward ratios for different uncle reward $\theta$ with $\lambda \tau(s) = 0.2$ ]{\includegraphics[width=2.0in,height=1.5in]{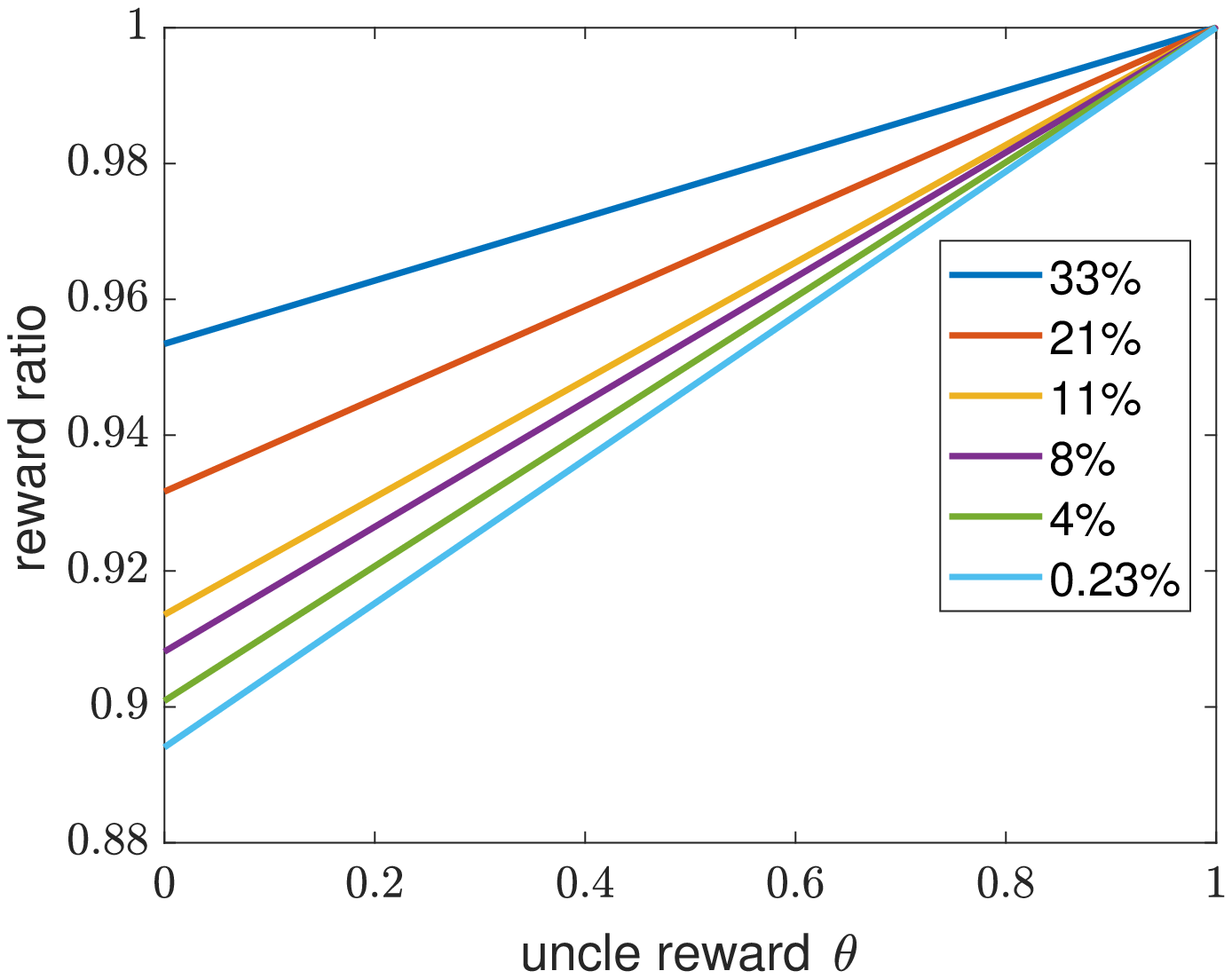}}
	\hfil
	\subfigure[Reward ratios for the growing hash rates with $\theta = 7/8$ and $\lambda \tau(s) = 0.2$ ]{\includegraphics[width=2.0in,height=1.5in]{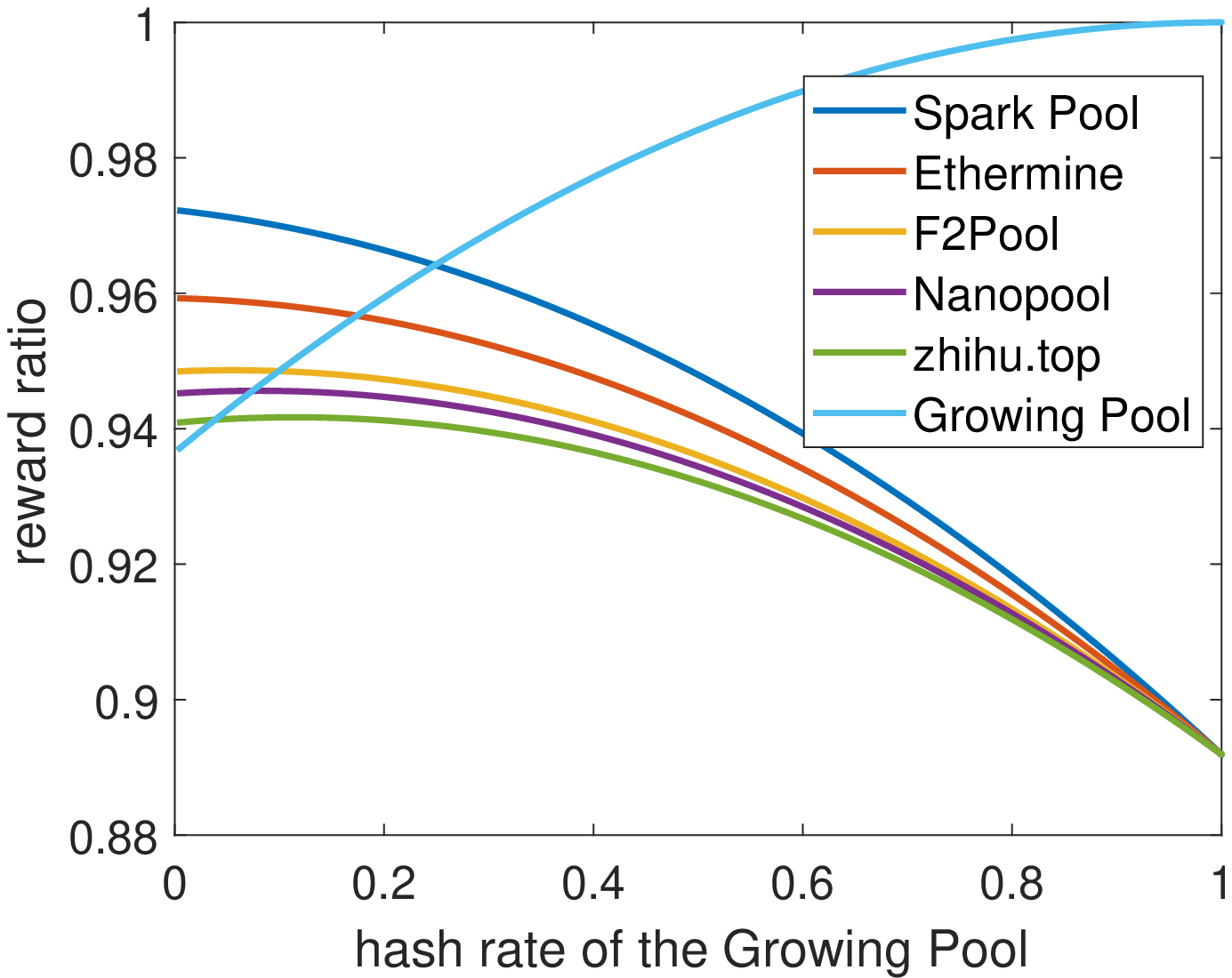}}
	\caption{Reward ratios for mining pools}
	\label{reward_analysis}
\end{figure*}

To explore the impact of temporary forks on the fairness of the mining
reward, we use the current mining pool hash rate to carry out
numerical experiments. Figure \ref{hash_rate_pie} presents the
hash rate fractions of mining pools in Ethereum
\cite{ethscan}. Note that even though Ethereum is not implementing the
longest chain rule, it provides a good source of data in large scale
on the distribution of generated blocks with shorter generation
intervals. The largest mining pool has dominated approximately 33\% of
the total hash rate. The top two mining pools have dominated over 50\%
of the total hash
rate. 
Accordingly, we assume that there are 5 large mining pools in the
simulation, and the ratio of the hash rate accounts for 33\%, 21\%,
11\%, 8\%, 4\%, and the remaining 23\% of the computing power is
evenly distributed by 100 small mining pools, each small mining pool
accounts for 0.23\% of the total hash rate.

We next investigate the fairness of the mining game. We first introduce the equity theory \cite{carrell1978equity}, which has been widely applied to business settings by social psychologists to describe the relationship between an employee's motivation and his or her perception of equitable or inequitable treatment. The fairness (or equity) is measured by comparing the ratio of benefits (or rewards) and contributions (or costs) for each person \cite{adams1965inequity}. According to the equity theory, partners do not have to receive equal benefits or make equal contributions, as long as the ratio between these benefits and contributions is similar, which will cause a feeling of satisfaction and help to have better outcomes.

Similar to the social business, fairness is also very important in the mining game. If the mining game is not fair, i.e., some of the mining pools are at advantage with higher ratio of rewards and contributions, while some of the mining pools are at disadvantage with lower ratio of rewards and contributions, then miners in the disadvantaged mining pools are incentive to join the advantaged mining pools to achieve higher ratio of rewards and contributions, which will increase the degree of centralization of the blockchain system. 
To measure the fairness of the mining game, the reward ratio of the mining pool $i$ is defined as the normalized ratio of its expected mining reward and its hash rate fraction, which is given as follows:
\begin{equation}\nonumber
\text{Reward ratio}(i) = \frac{Y_i}{x_i R} = R\left( 1 - (1 - \theta) P^{\text{uncle}}_i  \right).
\end{equation}
Without temporary forks, the reward ratio of any mining pool should be the same, which is one of the properties that guarantee the decentralization of the blockchain system. However, after considering the temporary fork, this fairness may be broken. Figure \ref{reward_analysis} (a) shows that as the network delay increases, mining pools with large hash rates will benefit from a higher reward ratio, while mining pools with small hash rates will be at a disadvantage. Thus, we have the following insight.
\begin{insight}
	The temporary fork cased by network delay positively impacts the reward ratio of mining pools with large hash rates.
\end{insight}
Figure \ref{reward_analysis} (b) shows the impact of the uncle block reward on the reward ratio of the mining pools. It can be seen that as the proportion of uncle reward increases, the gap between the reward ratio between the mining pools decreases, and the fairness of the blockchain system will improve. Therefore, we have
\begin{insight}
	A higher uncle reward greatly reduces the impact of temporary fork and leads to more fairness between the mining pools.
\end{insight}
Figure \ref{reward_analysis} (c) illustrates the impact of the computing power growth of a small mining pool (a mining pool that initially accounted for 0.23\% of the computing power), from which we can obtain the following insight. 
\begin{insight}
	The mining pools in the blockchain system are in a competitive relationship. The computing power of one mining pool directly affects the reward of others.
\end{insight}

\begin{figure*}[!htb]
	\centering
	\subfigure[Uncle rates of miners for blocks 1-1920000]{
		\includegraphics[width=2in,height=1.4in]{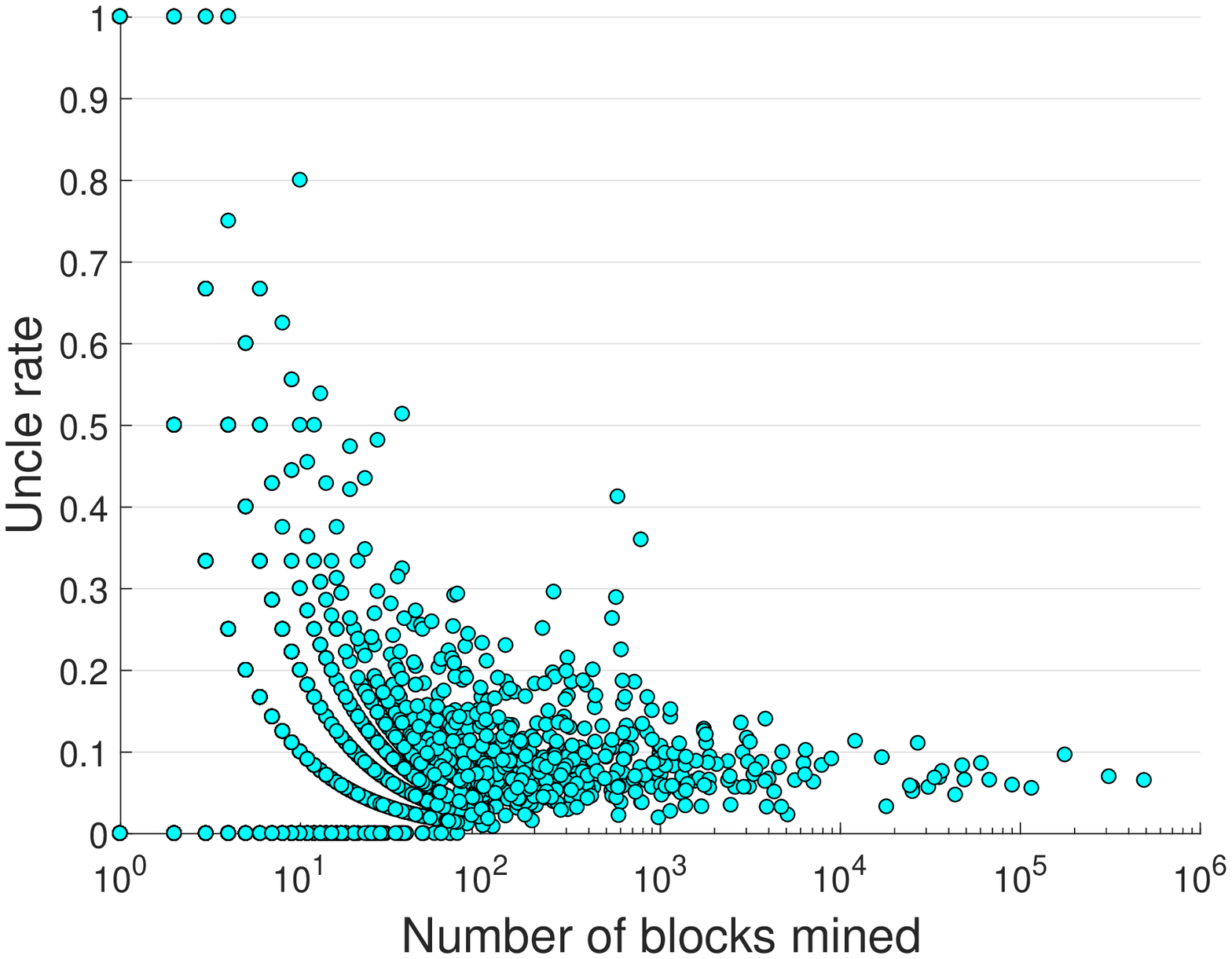}}
	\hfil
	\centering
	\subfigure[Fork rates of miners for blocks 1-1920000 ]{\includegraphics[width=2.0in,height=1.4in]{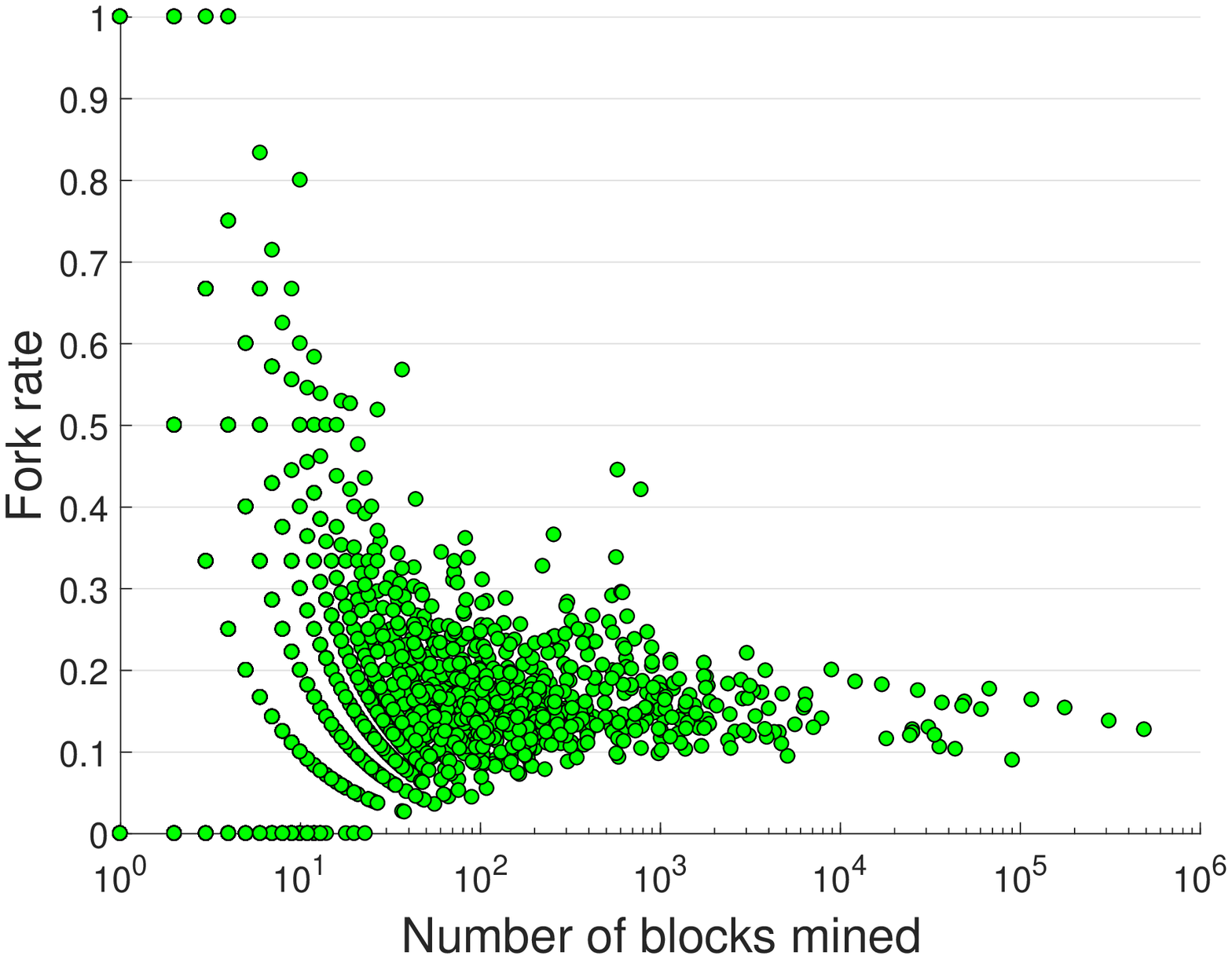}}
	\hfil
	\subfigure[Fail rates of miners for blocks 1-1920000 ]{\includegraphics[width=2.0in,height=1.4in]{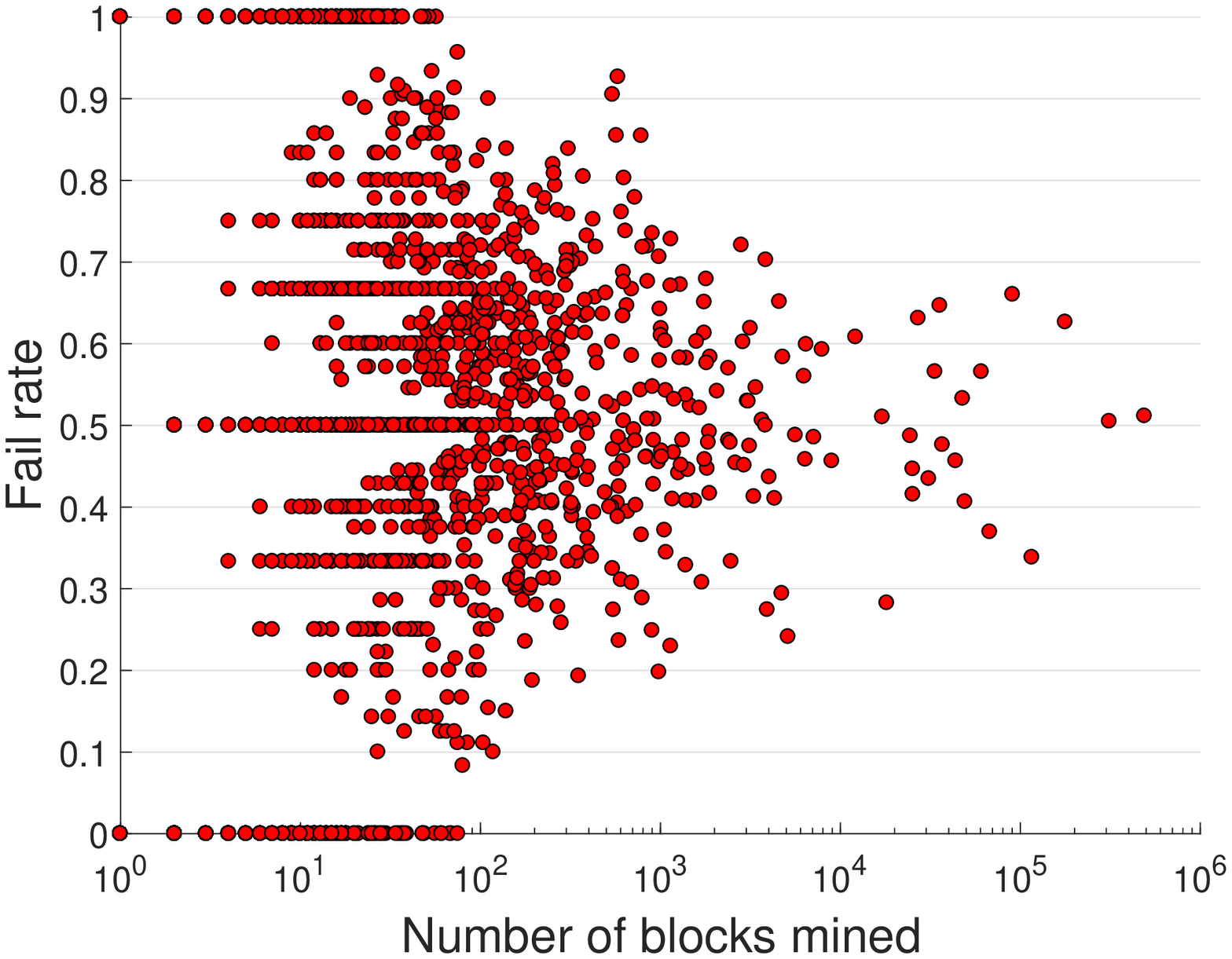}}
	\caption{Block statistics in Ethereum}
	\label{uncle_block}
\end{figure*}

To verify the correctness of our previous analysis and demonstrate the impact of temporary fork on the reward of mining pools in reality, we collect and analyze data for the first 1,920,000 blocks of the Ethereum blockchain (before Ethereum hard fork). 
Figure \ref{uncle_block} and Table \ref{statistic_table} show that there is a general trend that larger miners or mining pools (with more blocks mined) have lower fork rates and lower fail rates when temporary fork happens, thus have lower uncle rates and higher mining reward. Therefore, the miners or mining pools with large hash rates can obtain skew mining reward, while those with small hash rates will be trapped at disadvantage in the mining process, which is consistent with our previous analysis. 

\begin{table}[!htb]
	\caption{Block Statistics in Ethereum Segregating by Mining Pool Sizes}
	\centering
	\begin{tabular}{@{}llll@{}}
		\toprule
		Number of blocks mined       &  Uncle rate &  Fork rate &  Fail rate \\ \midrule
		(0, 10{]}                    & 0.111250           & 0.177345          & 0.595026          \\
		(10, 100{]}                  & 0.100081           & 0.173927          & 0.553244          \\
		(100, 1000{]}                & 0.091034           & 0.162981          & 0.533467          \\
		(1000, 10000{]}              & 0.077994           & 0.150922          & 0.503296          \\
		(10000, $\infty$) & 0.070361           & 0.141287          & 0.498919          \\ \bottomrule
	\end{tabular}
	\label{statistic_table}
\end{table}

Besides the analytical analysis on the temporary fork, here are also some other hypotheses which may help to explain the fact that larger miners have lower uncle rates \cite{more_uncle_statistics}, for example, large mining pools are professional operations and have more resources available to invest in improving their overall connectivity to the network and thus have higher efficiency \cite{lewenberg2015bitcoin}.

\section{Evolutionary Game for Mining Pool Evolution Equilibrium Analysis}\label{game}
Based on the temporal fork modeling analysis above, we then further consider studying the evolution equilibrium (i.e., the converging states) of the competing mining pools when the large population of miners can adapt their pool selections dynamically.
And the evolutionary game theory \cite{evolution} is an ideal modeling tool to analyze their evolution equilibrium due to the dynamic competitive behaviors of the miners.


\subsection{Evolutionary Game Model}

Generally, a miner is required to provide a certain hash rate of $\omega_{i}$ to join the mining pool $i$. 
Let $\bm{\omega} = [\omega_{1}, \cdots, \omega_{M}]^T$ denote the vector of hash rate specifications of the mining pools. Meanwhile, let $\bm{r} = [r_1, \cdots, r_M]^T$ denote population fraction vector for the mining pools, i.e., $ \bm{r} \in \Delta = \{\mathbf{r} | 0 \leq r_i \leq 1,\forall i \in \{1,\cdots,M\} \text{ and } \sum_{i=1}^{M} r_{i}=1 \} $. 
For example, if 20\% of miners join the mining pool, then $r_i = 0.2$. Therefore, for a blockchain with $N$ miners and $M$ mining pools, the hash rate of the mining pool $i$ is $h_i = N r_i \omega_i$, and the total hash rate of the entire network is $V = \sum_{j=1}^{M} N r_j \omega_j$. Thus, we have 
\begin{equation}\nonumber
	x_i = \frac{h_i}{V} = \frac{N r_i \omega_i}{\sum_{j=1}^{M} N r_j \omega_j} = \frac{\omega_i r_i}{\sum_{j=1}^{M} r_j \omega_j}.
\end{equation}
To characterize the evolution equilibrium of the mining pools in competition, we actually would like to derive the convergent population state of the miner distributions, which is useful to reveal the long-term trends on the degree of centralization of computing power distributions.

For simplicity, we consider the homogeneous mining cost and $p$ denotes the unit cost (e.g., expenditure for mining energy). The total cost for a miner in the mining pool $i$ can be expressed as $p \omega_{i}$.
In the blockchain system, there are many ways for a mining pool to distribute its mining reward, such as PPS (Pay Per Share), PPLNS (Pay Per Last N Shares). To simplify the model, we consider that the mining reward of the mining pool is allocated according to the proportion of miners' hash rates in the mining pool. Then the expected payoff of a miner in pool $i$ can be expressed as 
\begin{small}
\begin{eqnarray}\label{game_reward}
\begin{aligned}
&y_i(\bm{r}) = \frac{Y_i}{N r_i} - p\omega_{i} \\
&= \frac{R x_i}{N r_i} \! \left( 1 \! - \! (1 - \theta) P^{\text{uncle}}_i \right) - p\omega_{i}\\
&= \frac{R x_i}{N r_i} \! \left(\!\! 1 \! - \! (1 \!-\! \theta) \frac{1 - e^{-\lambda \tau(s)}}{2} \sum_{j \neq i}{x_j (1 - x_i + x_j)} \! \right) \!\! - p \omega_{i} .
\end{aligned}
\end{eqnarray}
\end{small}
Following the evolutionary game theory, we use the replicator dynamics to express the evolution dynamics, which can approximate the dynamic competitive mining pool selection behaviors of the miners from the whole population perspective \cite{evolution_pool}. Mathematically, the replicator dynamics of mining pool population changing can be expressed as the following ordinary differential equations, i.e.,
	\begin{equation}\label{ode}
	\dot{r}_i(t) \!=\! f_i(\bm{r}(t)) \!=\! r_i(t)\left(y_i(\bm{r}(t)) - \bar{y}(\bm{r}(t))\right),
	\end{equation}
where $\dot{r_i}(t)\triangleq\frac{dr(t)}{dt}$ is the growth rate of the mining pool $i$, and $\bar{y}(\bm{r}) = \sum_{i=1}^{M}{y_i(\bm{r})r_i}$ denotes the average payoff of all the miners in the blockchain system. Let $\mathbf{y}(\mathbf{r})=\left[y_{1}(\mathbf{r}), \ldots, y_{M}(\mathbf{r})\right]^{\top}$ denote payoff vector for all the mining pools at time $t$. Intuitively, replicator dynamics in (\ref{ode}) can capture the fact that the miner population of a mining pool that outperforms the average will increase (with the growing rate proportional to its advantage), while the population of a poorly-performed mining poor gradually decreases. 

We then define the evolution equilibrium of the mining pool evolutionary game above.
Suppose that there exists another population state $\mathbf{r}^{'}$ trying to ``invade'' (or substitute) the state $\mathbf{r}^{*}$ as the equilibrium state by attracting a small share $\epsilon \in (0, 1)$ in the population of miners to switch to $\mathbf{r}^{'}$. Then, $\mathbf{r}^*$ is an Evolution Stable Strategy (ESS) if the following condition holds for all $\epsilon \in (0, \bar{\epsilon})$ \cite{evolution_pool,webb2007game}:
\begin{equation}\label{eq_ESS}
\sum_{i=1}^{M} r_{i}^{*} y_{i}\left((1-\epsilon) \mathbf{r}^{*}+\epsilon \mathbf{r}^{\prime}\right) > \sum_{i=1}^{M} r_{i}^{\prime} y_{i}\left((1-\epsilon) \mathbf{r}^{*}+\epsilon \mathbf{r}^{\prime}\right) .
\end{equation}
Indeed, ESS does not necessarily exist. It is useful to look for neutrally stable strategies when there is no ESS \cite{van2010but}. $\mathbf{r}^*$ is a Neutrally Stable Strategy (NSS) if the following condition holds for all $\epsilon \in (0, \bar{\epsilon})$
\begin{equation}\label{eq_NSS}
\sum_{i=1}^{M} r_{i}^{*} y_{i}\left((1-\epsilon) \mathbf{r}^{*}+\epsilon \mathbf{r}^{\prime}\right) \geq \sum_{i=1}^{M} r_{i}^{\prime} y_{i}\left((1-\epsilon) \mathbf{r}^{*}+\epsilon \mathbf{r}^{\prime}\right) .
\end{equation}

Intuitively, the ESS (NSS) presents the convergent stable states for the mining pool evolution dynamics. 
ESS (NSS) is a best reply to itself and a (weakly) better reply to all other best replies than these are to themselves \cite{banerjee2000neutrally}. Furthermore, ESS implies asymptotic stability in the replicator dynamics \cite{taylor1978evolutionary} and NSS implies Lyapunov stability \cite{thomas1985evolutionarily, weibull1997evolutionary}. The evolution equilibrium is said to be Lyapunov stable if no small disturbances can bring it to move far away, and asymptotic stability implies stronger stability such that the population will eventually return to the equilibrium when experiencing small disturbances.
Therefore, an ESS (NSS) has the property that the miners at the equilibrium can achieve the stable mutually-satisfactory states without incentives to deviate and meanwhile is robust to small perturbations by some miners' random or irrational deviating behaviors.

\subsection{Equilibrium Analysis}\label{equal}

We first study the case of $M$ mining pools with the same hash rate specifications for each miner, i.e., $\omega_{1} = \cdots = \omega_{M} = \omega$. We can obtain Theorem \ref{same_w} as follows.

\begin{theorem}\label{same_w}
	For the mining pools with the same hash rate specifications, i.e., $\omega_{1} = \cdots = \omega_{M} = \omega$, the ESSs of the game always exist, and the ESSs of the game are $\mathbf{r} = \mathbf{e}_i, i \in \mathcal{M} = \{1, \cdots, M\}$, where $\mathbf{e}_i$ is a vector whose $i$-th component is 1, and the remaining components are 0, i.e., $\mathbf{e}_i = [0 \cdots 1 \cdots 0]^T$.
\end{theorem}
The proof is given in Appendix C.
Theorem \ref{same_w} shows that a single hash rate specification of all the pools will lead to the centralization trend of the miner population in the blockchain system. Therefore, we have
\begin{insight}
	The diversity of hash rates (e.g., with different mining equipment, such as CPU, GPU, FPGA, ASIC, etc.) in the blockchain system is more conducive to decentralization.
\end{insight}


We next consider the case of two mining pools with unequal hash rate specifications, and denote the equilibrium population states of these two pools as $(r^*,1-r^*)$ for simplicity. We obtain the results as follows.
\begin{theorem}\label{two_pool_unequal}
	For the mining game of two mining pools with unequal hash rate specifications $\omega_{1}, \omega_{2}$ ($\omega_{1} > \omega_{2} > 0$), the ESS of the game always exists.
	
	(1) The ESS of the game is $r^* = 1$, when 
	\begin{equation}\label{r_equal_1}
	p N \omega_{1} < R + \frac{R \left( 1 - e^{-\lambda \tau(s)} \right) \omega_{2} (1 - \theta)}{\omega_{1} - \omega_{2}}.
	\end{equation}
	
	(2) The ESS of the game is $r^* = 0$, when
	\begin{equation}\label{r_equal_0}
	p N \omega_{2} > R - \frac{R \left( 1 - e^{-\lambda \tau(s)} \right) \omega_{1} (1 - \theta)}{\omega_{1} - \omega_{2}}.
	\end{equation}
	
	(3) When 
	\begin{eqnarray}\nonumber
	\left\{
	\begin{array}{ll}
	p N \omega_{1} \geq R + \frac{R \left( 1 - e^{-\lambda \tau(s)} \right) \omega_{2} (1 - \theta)}{\omega_{1} - \omega_{2}},\\
	p N \omega_{2} \leq R - \frac{R \left( 1 - e^{-\lambda \tau(s)} \right) \omega_{1} (1 - \theta)}{\omega_{1} - \omega_{2}},
	\end{array}
	\right.
	\end{eqnarray}
	the ESS of the game $r*$  can be obtained by solving the following cubic equation with the constraint that $r \in (0, 1)$
	\begin{equation}\nonumber
	a r^3 + b r^2 + c r + d = 0 ,
	\end{equation}
	where 
	\begin{small}
	\begin{eqnarray}\nonumber
	\left\{
	\begin{array}{ll}
	\!\! a = N p (\omega_{1} - \omega_{2})^4, \\
	\!\! b = R (\omega_{1} - \omega_{2})^3 + R (1 \!-\! e^{-\lambda \tau(s)}) (1\!-\!\theta) \omega_{1} \omega_{2} (\omega_{1} - \omega_{2})\\
	\qquad - 3Np\omega_{2}(\omega_{1} - \omega_{2})^3 , \\
	\!\! c = 2R\omega_{2}(\omega_{1} - \omega_{2})^2 + 2 R (1 - e^{-\lambda \tau(s)}) (1-\theta) \omega_{1} \omega_{2}^2 \\
	\qquad- 3N p \omega_{2}^2 (\omega_{1} - \omega_{2})^2, \\
	\!\! d = \!-\omega_{2}^2 (R (1 \!-\! e^{-\! \lambda \tau(s)}) (1\!-\!\theta) \omega_{1} \!-\! (R \!-\! N p \omega_{2}) (\omega_{1} \!-\! \omega_{2}).
	\end{array}
	\right.
	\end{eqnarray}
	\end{small}
\end{theorem}
The proof is given in Appendix D. From Theorem \ref{two_pool_unequal}, we find that the mining pool with a larger hash rate specification will dominate the blockchain system, when the mining reward $R$ is large enough or the hash rate specifications of the mining pools are close enough (i.e., $\omega_{1} - \omega_{2}$ small enough). Second, we also see that given a fixed mining reward $R$, reducing the block generation speed $\lambda$ or the network propagation delay $\tau(s)$ can help to reduce the possibility that only one mining pool dominates at the equilibrium (i.e., cases (1) and (2) in Theorem \ref{two_pool_unequal}). This would help to maintain the degree of decentralization in blockchain system. Last but not least, such observation also holds when increasing the uncle reward ratio $\theta$. We further derive the closed-form result for the case that uncle reward ratio is sufficiently large (i.e., $\theta\rightarrow 1$) in Theorem \ref{uncle_reward_limit_two}.

\begin{theorem}\label{uncle_reward_limit_two}
	For the mining game of two mining pools with unequal hash rate specifications $\omega_{1}, \omega_{2}$ ($\omega_{1} > \omega_{2}$), if the uncle reward $\theta \rightarrow 1$, the ESS of the game always exists.
	
	(1) The ESS of the game is $r^* = 1$, if $R \geq p N \omega_{1}$.
	
	(2) The ESS of the game is $r^* = 0$, if $R \leq p N \omega_{2}$.
	
	(3) The ESS of the game is $r^* = \frac{R - \omega_{2} p N}{p N (\omega_{1} - \omega_{2})}$ if $p N \omega_{2} < R < p N \omega_{1}$.
\end{theorem}
The proof is given in Appendix E. From Theorem \ref{uncle_reward_limit_two}, we find that when the uncle reward is large enough, under the condition that the mining reward  $R$ is modest, i.e., $p N \omega_{2} < R < p N \omega_{1}$, then none of the mining pool will be able to dominate the system.



For a more general case with multiple pools, we can show the following results.

\begin{theorem}\label{general_case_limit}
	For the mining game of $M$ mining pools with unequal hash rate specifications, namely, $\omega_{1}, \cdots, \omega_{M}$ ($\omega_{1} > \omega_{2} > \cdots > \omega_{M}$), if the network propagation time is negligible, i.e., $\tau(s) = 0$, or the uncle reward is large enough, i.e., $\theta  = 1$, the NSSs of the game always exist. 
	
	(1) When $R \geq p N \omega_{1}$, the ESS of the game is $r^*=[r_1 \ \cdots \ r_M]$, where $r_1 = 1$ and $r_j = 0, \forall j \in \{2, \cdots, M\}$.
	
	(2) When $R \leq p N \omega_{M}$, the ESS of the game is $r^*=[r_1 \ \cdots \ r_M]$, where $r_M = 1$ and $r_j = 0, \forall j \in \{1, \cdots, M-1\}$.
	
	(3) When $p N \omega_{M} < R < p N \omega_{1}$, $\Delta^{\text{NSS}} = \{ \mathbf{r}^* | \sum_{i=1}^{M}{r_i^* \omega_{i} = \frac{R}{p N}}, \mathbf{r}^* \in \Delta  \}$ is the set of NSSs of the game. Moreover, $\{ \mathbf{r}^* | r_i^* > 0, \forall i \in \{1,\cdots,M\}\text{ and } r^* \in \Delta^{\text{NSS}} \}$ are the asymptotically stable states of the replicator dynamics system.
\end{theorem}
The proof is given in Appendix F.
Theorem \ref{general_case_limit} reveals the convergent stable states for the mining pool evolution dynamics when the uncle reward is large enough (i.e., $\theta =1$).  We find that when the uncle reward is large enough, under the condition that the mining reward  $R$ is modest, i.e., $p N \omega_{M} < R < p N \omega_{1}$, then none of the mining pool will be able to dominate the system. Additionally, in this regime there are infinite NSSs and asymptotically stable states of the game, and the convergent equilibriums of mining pools will depend on the initial population states. 

Combining Theorem 3, Theorem 4 and Theorem 5, we have the following insight.
\begin{insight}
	The disproportionate mining reward caused by temporary fork positively impacts the degree of centralization of the blockchain network. While a higher uncle block reward helps to maintain the decentralization of the blockchain network.
\end{insight}

\subsection{Numerical Experiments}\label{numerical}

In this section, we conduct numerical simulations and evaluate the evolution equilibrium of the mining pools in different situations. We first consider a blockchain network with $N = 5000$ individual miners, which evolve to form two mining pools (i.e., $M = 2$), noted as $A, B$. For the purpose of demonstration, we set the block generation parameters as $\lambda = 1/10, R = 1200, p = 0.01$. We also set the initial population state as $r = [0.6;0.4]$. We first consider that the two pools adopt different computation power specifications, $\omega_{1} = 30, \omega_{2} = 20$.

\begin{figure*}
	\begin{minipage}{0.495\textwidth}
		\subfigure[mining without uncle block reward]{
			\includegraphics[width=1.7in,height=1.5in]{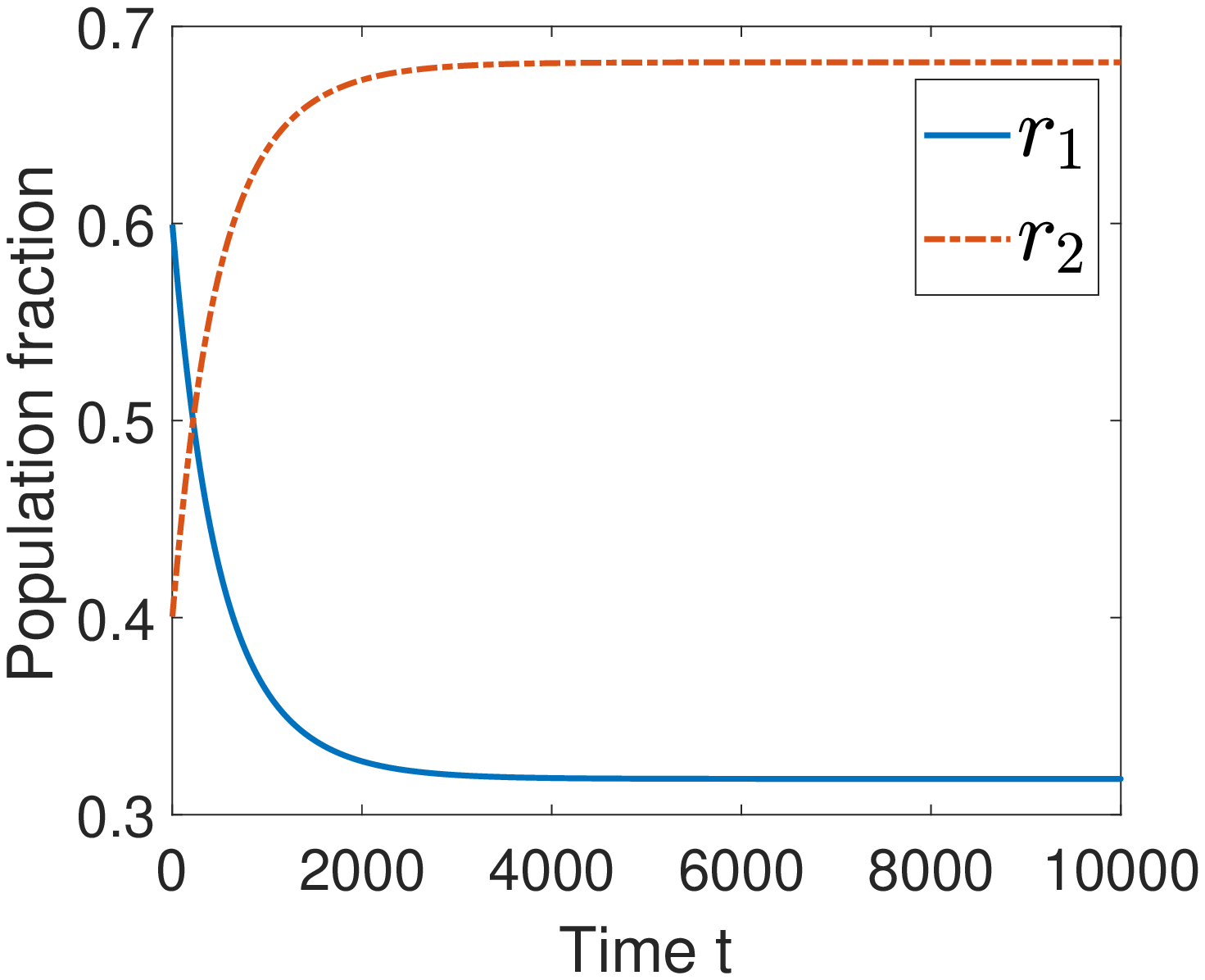}}
		\subfigure[mining with uncle block reward $\theta = 0.5$ ]{\includegraphics[width=1.7in,height=1.5in]{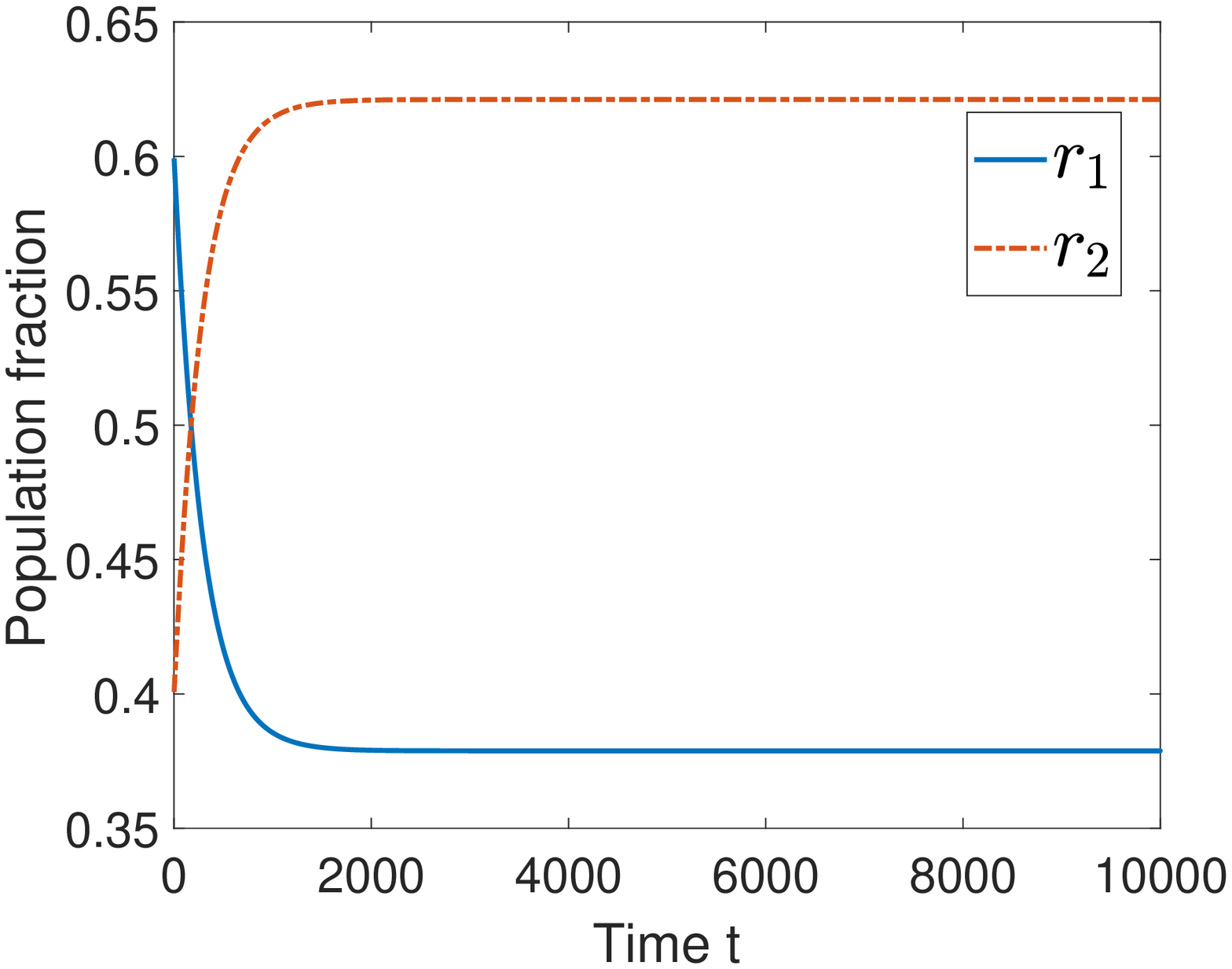}}
		\caption{Evolution of the miner's population states over time with $\quad$ propagation delay $\tau(s) = 0.5$}
		\label{tau_1}
	\end{minipage}
	\begin{minipage}{0.495\textwidth}
		\subfigure[mining without uncle block reward]{
			\includegraphics[width=1.7in,height=1.5in]{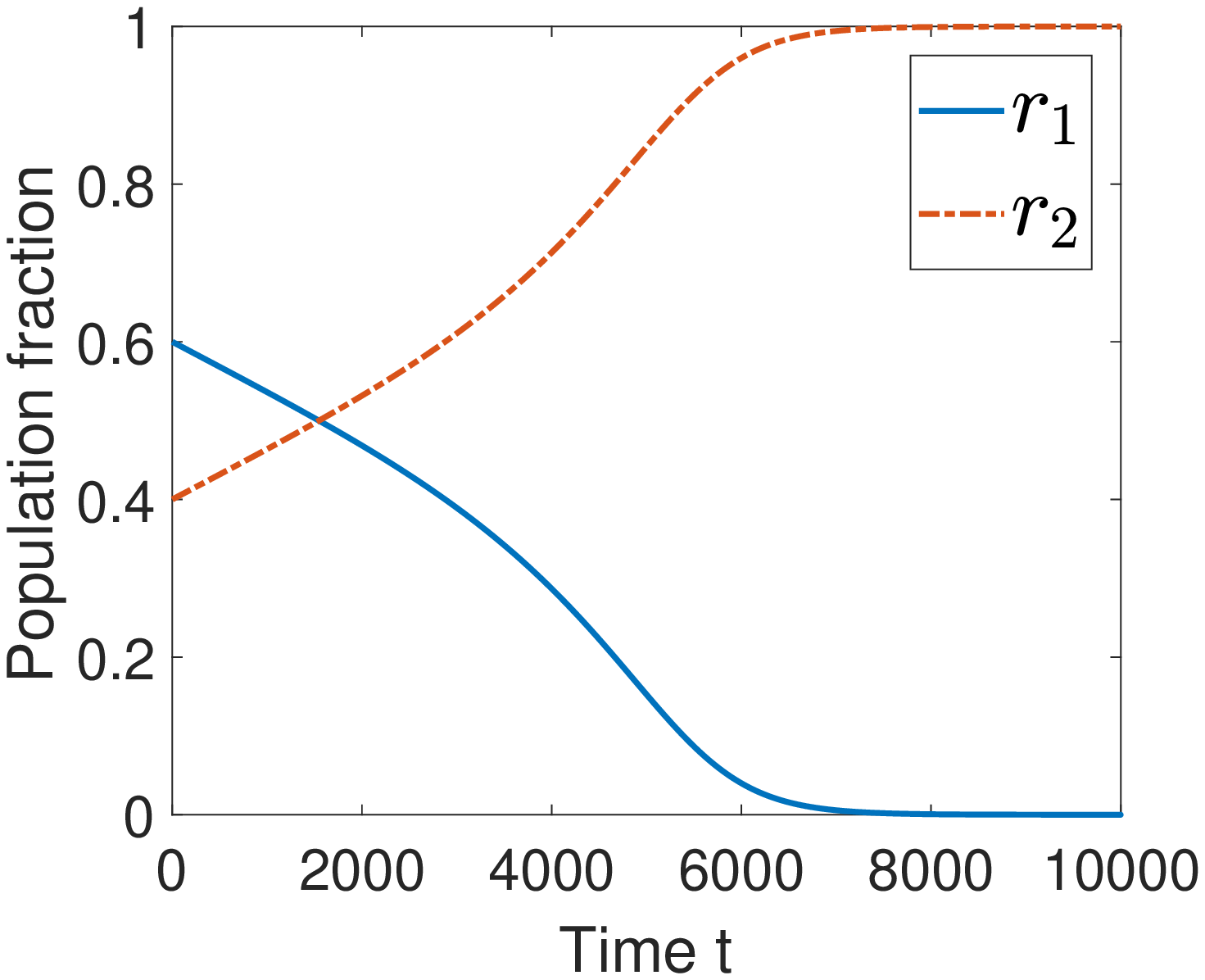}}
		\subfigure[mining with uncle block reward $\theta = 0.5$ ]{\includegraphics[width=1.7in,height=1.5in]{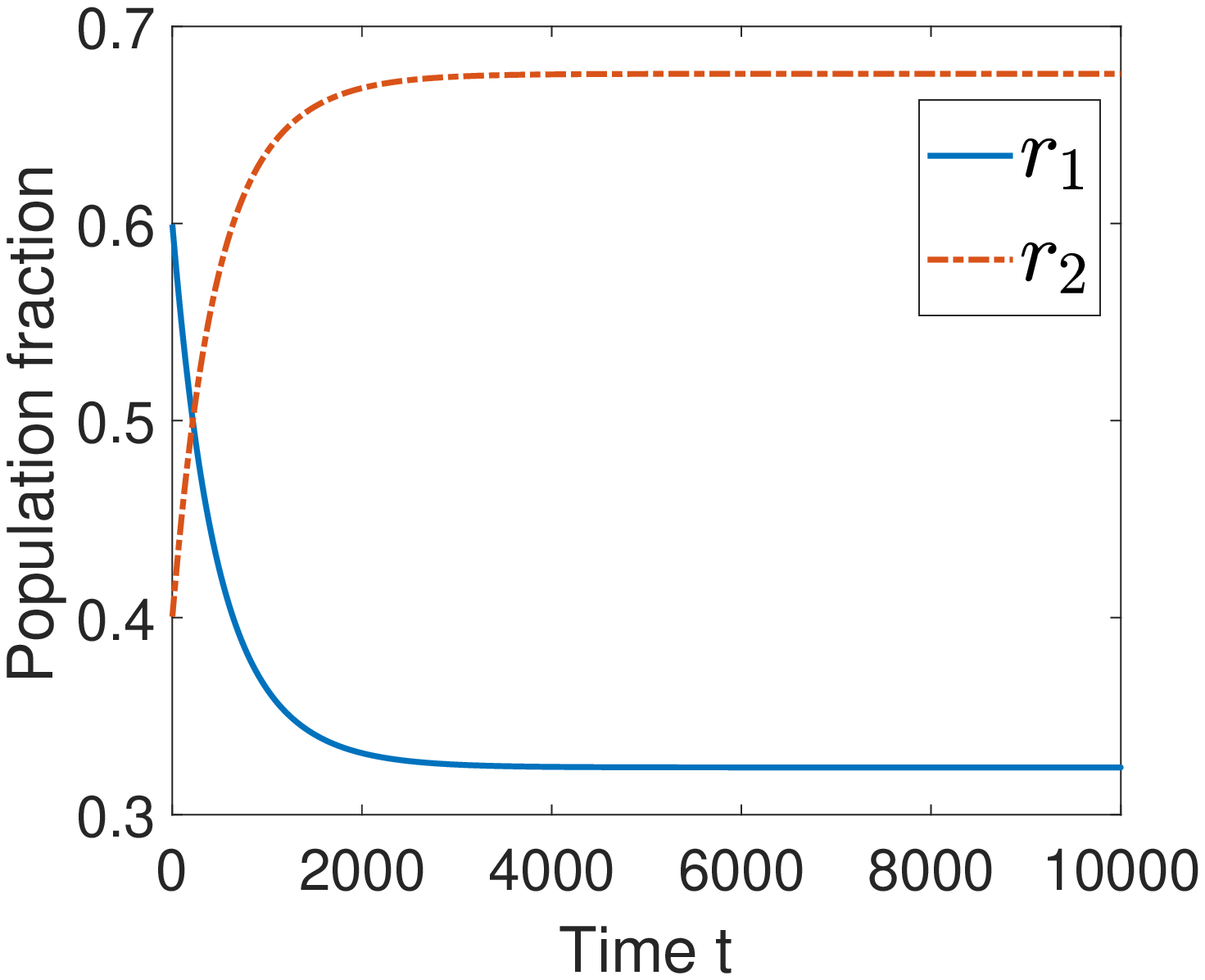}}
		\caption{Evolution of the miner's population states over time with propagation delay $\tau(s) = 1$}
		\label{tau_3}
	\end{minipage}
\end{figure*}

\begin{figure*}[!htb]
	\subfigure[mining without uncle block reward]{
		\includegraphics[width=1.7in,height=1.5in]{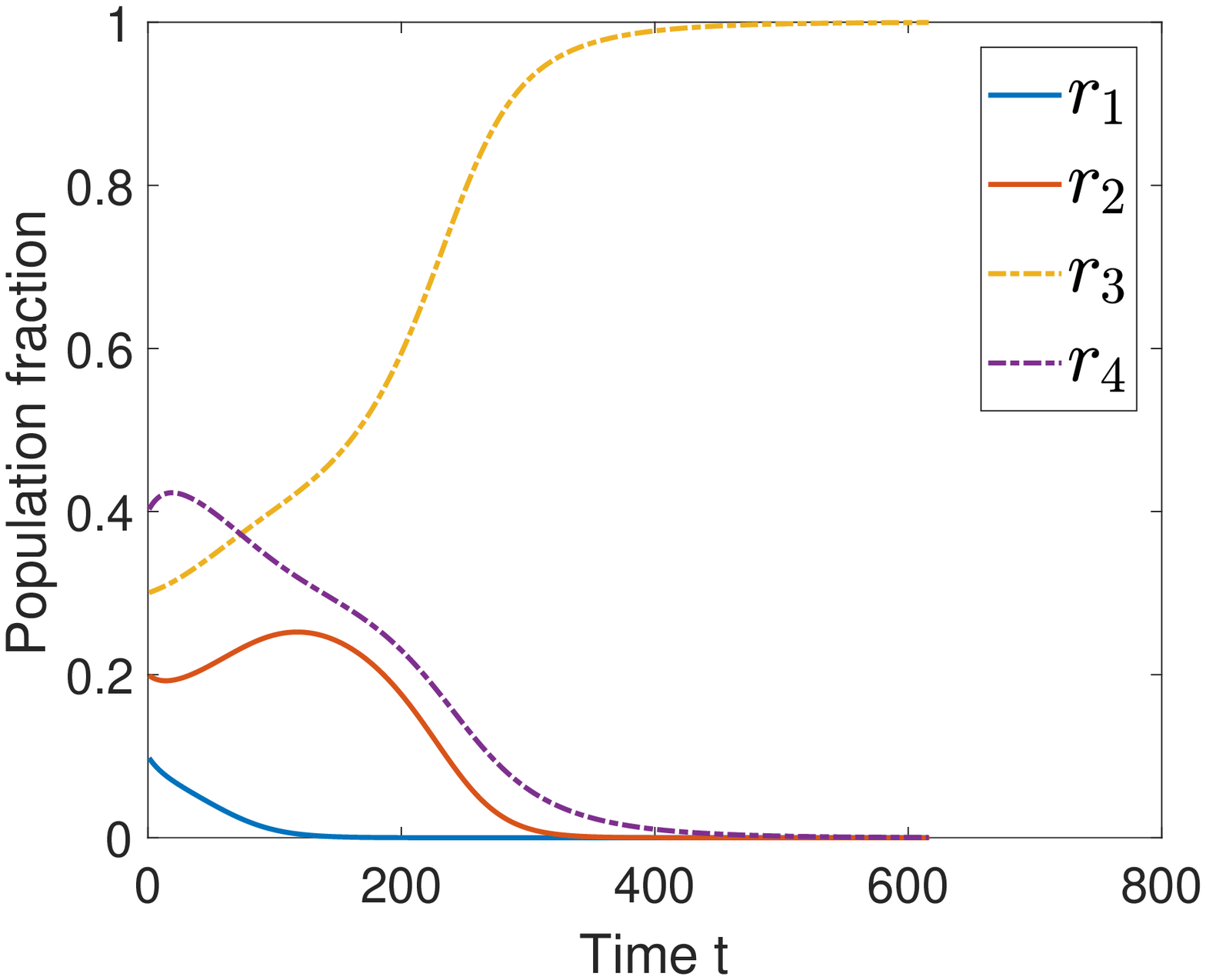}}
	\subfigure[mining with uncle block reward $\theta = 7/8$ ]{\includegraphics[width=1.7in,height=1.5in]{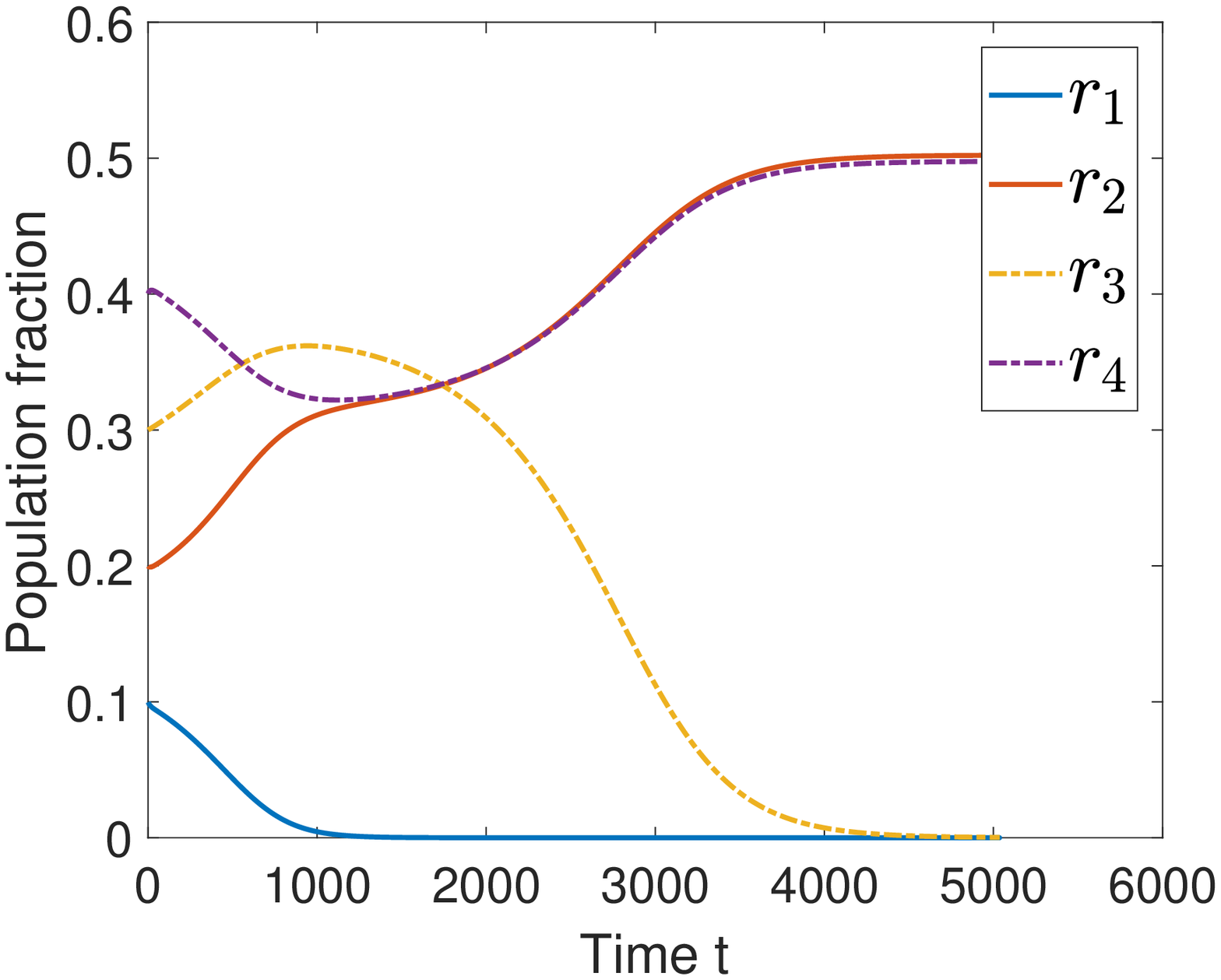}}
	\subfigure[mining with uncle block reward $\theta \rightarrow 1$
	]{\includegraphics[width=1.7in,height=1.5in]{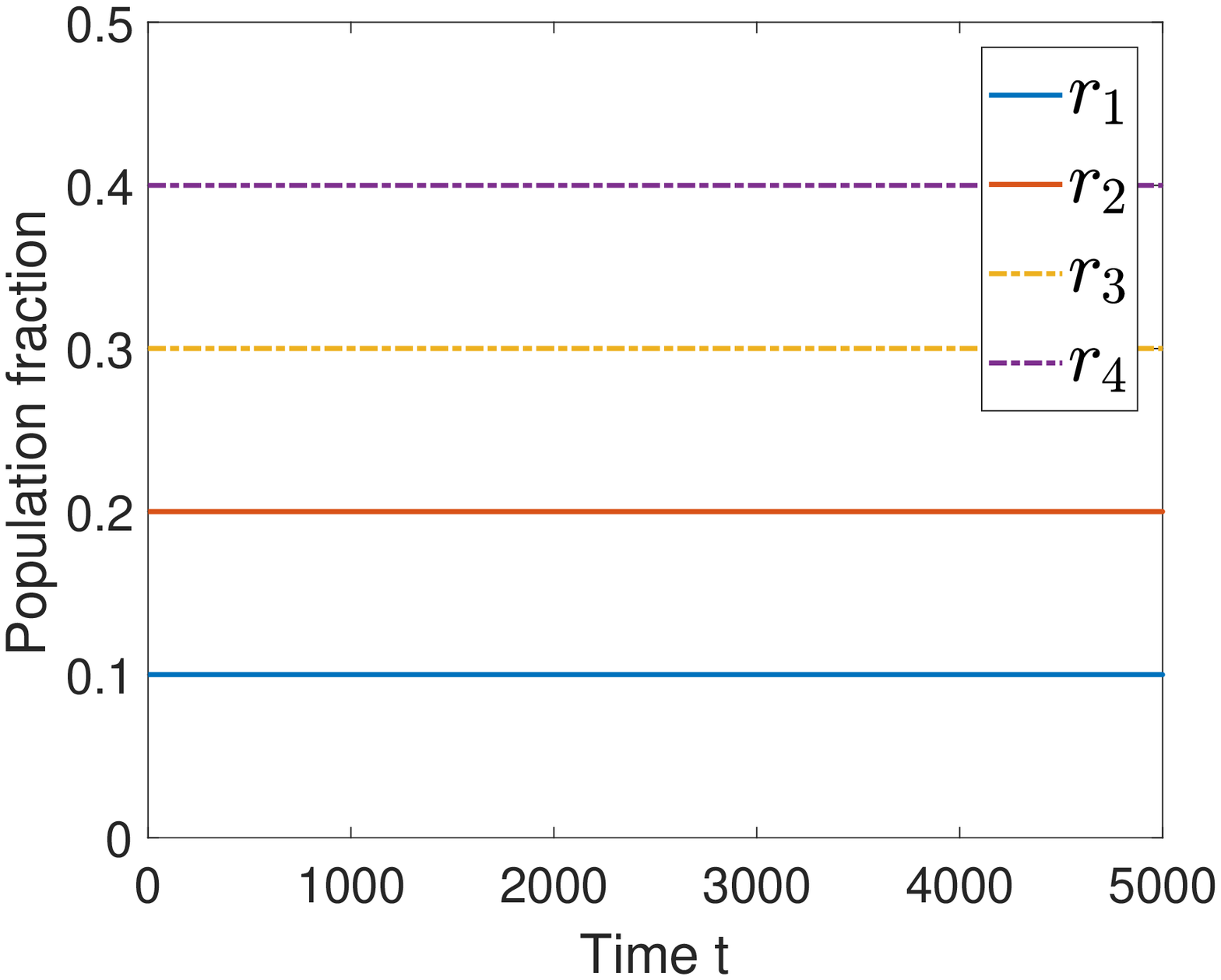}}
	\subfigure[mining power distribution when $\theta \rightarrow 1$
	]{\includegraphics[width=1.7in,height=1.5in]{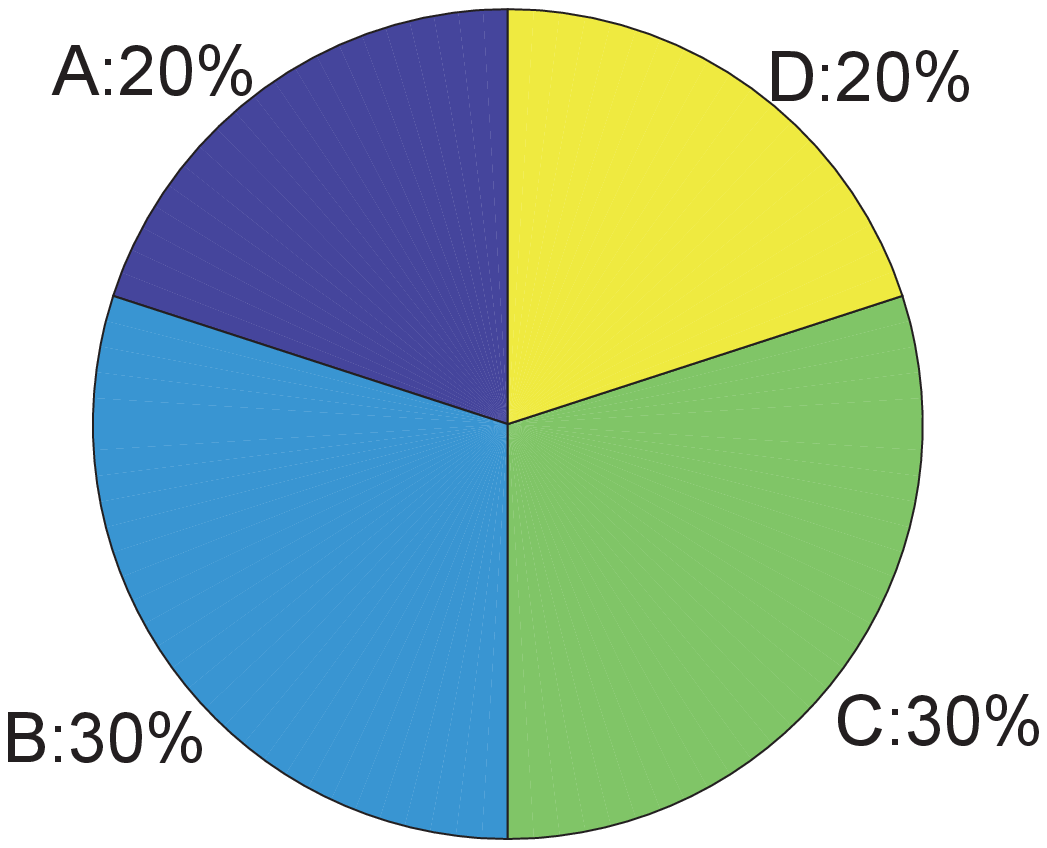}}
	\caption{Evolution of four mining pools}
	\label{4_pool}
\end{figure*}

\begin{figure*}[!t]
	\centering
	\includegraphics[width=5.5in,height=1.6in]{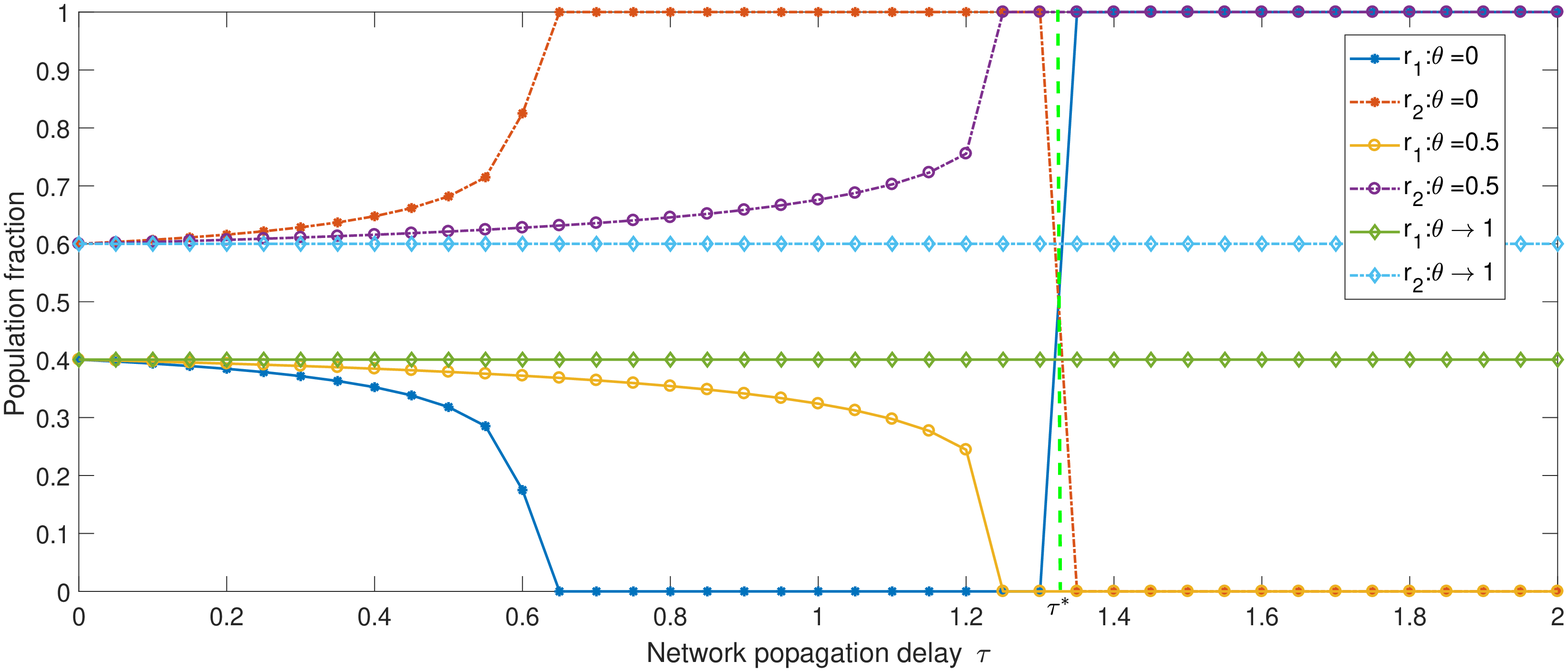}
	\caption{The convergent stable state with different network delay $\tau$ and uncle block reward $\theta$}
	\label{diff_tau}
\end{figure*}


Figure \ref{tau_1} demonstrates the evolution process of the mining pools driven by the replicator dynamics  with network propagation delay $\tau(s) = 0.5$. Figure \ref{tau_1} (a) shows that if there is no uncle block reward, i.e., $\theta = 0$, the number of miners in the mining pool $B$ will exceed more than 65\% of the total population. Figure \ref{tau_1} (b) shows that when considering uncle block reward, i.e., $\theta = 0.5$, the system will be less centralized. Figure \ref{tau_3} shows the similar results but with network propagation delay $\tau(s) = 1$. According to Figure \ref{tau_3} (a), we find that when network propagation delay is large, the degree of centralization of the blockchain system will be higher, but when increasing the uncle block reward, the system will be less centralized.


Additionally, we consider a more general situation with four mining pools, noted as $A, B, C, D$, where each pool adopts different requirements on the hash rate specifications as $\omega_{1} = 40, \omega_{2} = 30, \omega_{3} = 20, \omega_{4} = 10$ with initial population state as $\bm{r} = [0.1,0.2,0.3,0.4]$, evolves in the blockchain network with $\lambda = 1 / 10, \tau(s) = 1, R = 1500, N = 5000, p = 0.015$. The evolution of miner population states is presented in Figure \ref{4_pool}. In the Figure \ref{4_pool} (a), we observe that when the miners' pool-selection strategies converge to the equilibrium, the mining pool $B$ strictly dominates. But when adopting uncle reward with $\theta = 7 /8$, none of the mining pools can strictly dominate the blockchain network, which again shows that uncle block reward helps to impove the degree of decentralization of the blockchain network. Furthermore, when the uncle block reward is large enough, i.e., $\theta \rightarrow 1$, the mining pools will evolve to the convergent stable state shown in Figure \ref{4_pool} (c), and the mining power distribution at the stable state is shown in Figure \ref{4_pool} (d).



Finally, we analyze the influence of the network delay and uncle block reward on the evolution of the mining pools. Figure \ref{diff_tau} shows the equilibrium state of two competing pools with different network delay $\tau$ and uncle block reward $\theta$. We find that as the delay of the network propagation increases, the degree of centralization of the system will also increase until one mining pool will dominate the whole blockchain system. As shown in Figure \ref{diff_tau}, when $\tau$ is less than an observed threshold $\tau^*$, as the network propagation time $\tau$ increases, miners will tend to join the pool with a smaller hash rate specification. However, when the propagation delay is very large, i.e., $\tau > \tau^*$, the mining pool with a larger hash rate initially will dominate. This is mainly because a larger delay will lead to a higher probability of temporary fork, which will trap the mining pools with smaller hash rates at a disadvantage in the competition. In addition, Figure \ref{diff_tau} shows that the uncle block reward can greatly reduce the impact of temporary forks caused by network propagation delay. When the uncle block reward is large enough, i.e., $\theta \rightarrow 1$, the effect of network propagation delay is negligible, thus the decentralization of the blockchain system will be maintained.


\subsection{Distributions of Mining Power in Realistic Blockchain Systems}

\begin{figure}[t]
	\centering
	\subfigure[Distributions of mining power in MSR, ETH, BCH and BTC]{
		\includegraphics[width=1.66in,height=1.5in]{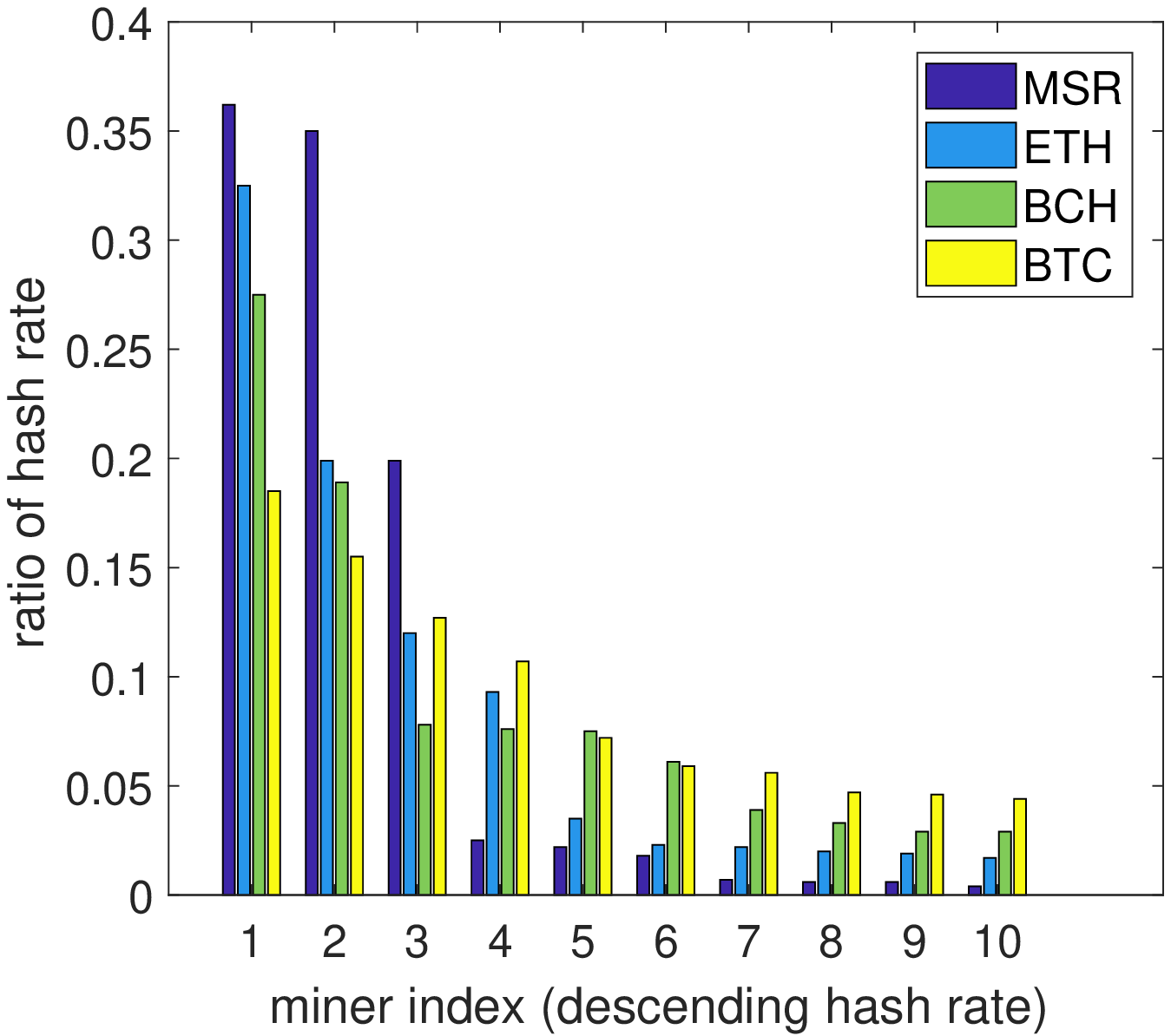}}
	\subfigure[Gini coefficients of different blockchain systems ]{\includegraphics[width=1.66in,height=1.5in]{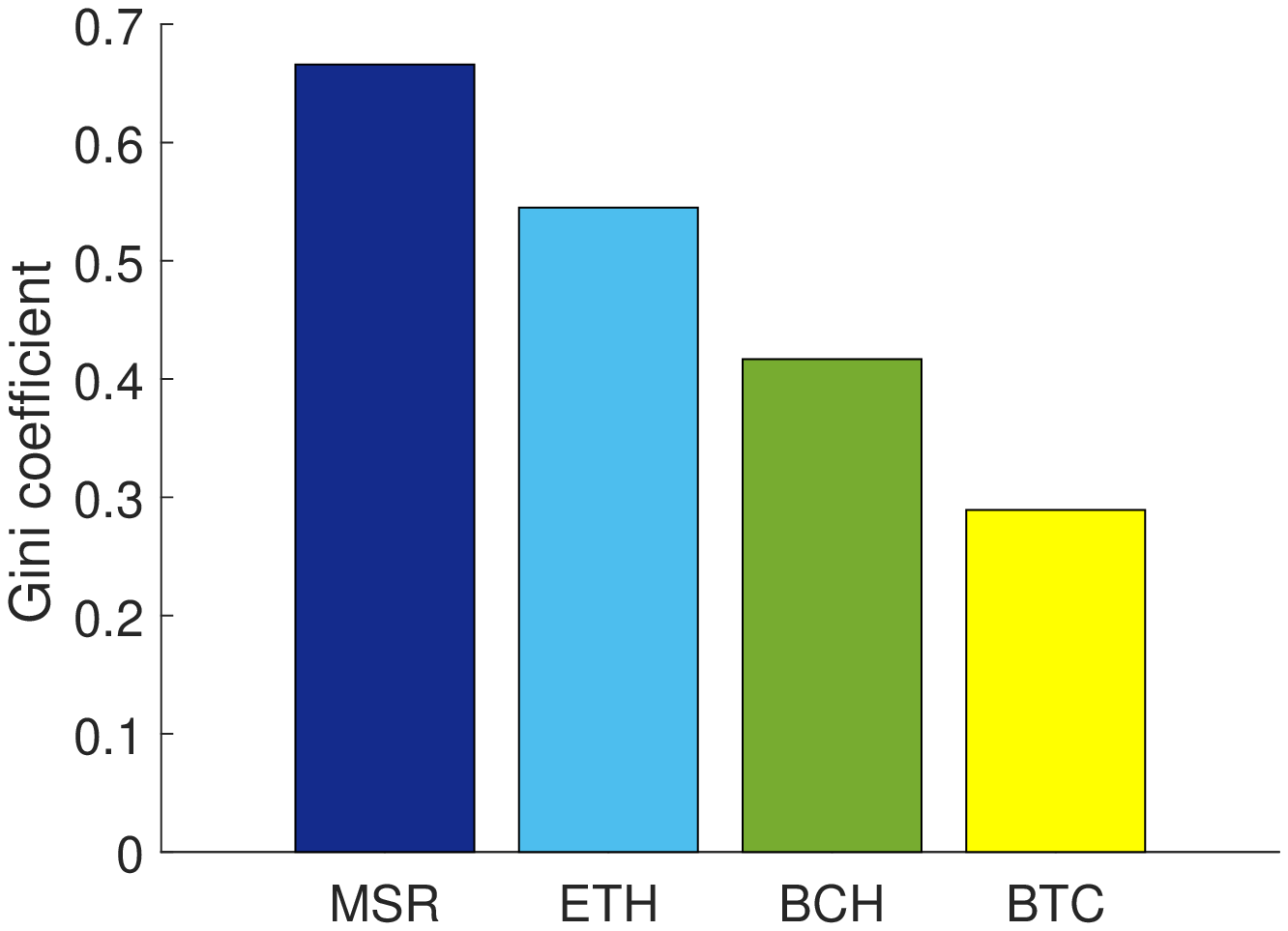}}
	\caption{Distributions of mining power in four major blockchains}
	\label{distribution}
\end{figure}

In this subsection, we analyze the distributions of computing power in existing typical blockchain systems, namely, Masari (MSR), Ethereum (ETH), Bitcoin (BTC) and Bitcoin Cash (BCH), which verifies the rationality of the previous theorems in this section.

Bitcoin and Ethereum are two of the leading cryptocurrencies with the largest market capitalization and user populations \cite{decentralization}. Bitcoin's block generation interval is about 10 minutes \cite{bitcoin}, while Ethereum's block generation interval is about 10-20 seconds \cite{ether}. A significant difference between Ethereum and Bitcoin is that Ethereum uses the GHOST protocol, which has uncle block reward. Bitcoin Cash (BCH) \cite{bitcoincash} is a new chain generated by Bitcoin due to the hard fork. One of the significant differences between BCH and BTC is that BCH expands its block size, and BCH can support a larger block size than that in BTC. Because the uncle block reward in Ethereum may lead to the behavior of uncle mining, Masari (MSR) uses the SECOR protocol \cite{uncle_mining} to avoid uncle mining by reducing uncle reward. Specifically, the uncle block reward in Ethereum is 7/8 of the basis block reward, while the uncle block reward in MSR is 1/2 of the basis block reward.

%

To better measure the degree of centralization of the blockchain, we introduce the Gini coefficient \cite{sen1997economic}, a single number aimed at measuring the degree of inequality in a distribution, which is widely used in economics to measure the inequality among values of a frequency distribution, e.g., levels of income. Based on the mathematical definition of the Gini coefficient, the degree of centralization of the blockchain with $M$ mining pools is given by:
$$
G = \frac{\sum_{i=1}^{M}\sum_{j=1}^{M} | x_i - x_j | }{2M \sum_{i=1}^{M} x_i } = \frac{\sum_{i=1}^{M}\sum_{j=1}^{M} | x_i - x_j | }{2M^2 \bar{x} },
$$
where $x_i$ denotes the normalized hash rate of the mining pool $i$.

The greater value of the Gini coefficient implies that the blockchain system is more centralized. The degree of centralization (the Gini coefficient) of the blockchain can theoretically range from 0 (complete equality or complete decentralization, i.e., the mining power of each mining pool is equal) to 1 (complete inequality or complete centralization, i.e., all the miners join one of the mining pools while the remaining pools are without any mining power). These properties greatly support that the Gini coefficient above can well measure the degree of centralization of the blockchain.

Figure \ref{distribution} (a) shows the distribution of hash rates in the top 10 mining pools (which occupy more than 90\% of the total computing power) in Masari (MSR), Ethereum (ETH), Bitcoin Cash (BCH) and Bitcoin (BTC), respectively. Furthermore, to better measure the degree of centralization in these blockchains, we calculate the Gini coefficient of different blockchain systems using the distribution of hash rates in the top 10 mining pools. As shown in Figure \ref{distribution} (b), the calculated Gini coefficients of MSR, ETH, BCH and BTC are correspondingly 0.67, 0.54, 0.42 and 0.29. We can find that the order of the degree of centralization of the blockchain systems from high to low is MSR > ETH > BCH > BTC.


According to our analysis above, increasing the uncle reward can increase the degree of decentralization of the system, and the uncle reward in MSR is smaller than that in ETH, so the centralization degree of MSR is greater than that in ETH. Although there are uncle rewards in ETH and MSR to reduce the impact of temporary forks, the block production speeds of ETH and MSR are much faster than that of BTC and BCH, leading to more significant advantages for large ETH and MSR mining pools in getting rewards. Therefore, ETH and MSR tend to be more centralized than BTC and BCH. Since the block size $s$ in BCH is larger than that in BTC, the network delay $\tau(s)$ of BCH is larger than that of BTC, thus the hash rate in BCH is more concentrated. These demonstrate that our models can provide useful insights for practical blockchain systems.

\section{Related Work}

There has been a rich body of previous work on blockchain network and mining pool. In the following, we introduce related work regarding temporary fork, game theory, the impact of uncle block rewards and the decentralization in blockchain networks, respectively.

\textbf{Temporary Fork in Blockchain Networks.} Due to the network propagation delay, blockchain networks may face inconsistencies in form of temporary forks. There are several papers studying the delay and temporary fork in blockchain networks from a networking perspective. In \cite{network_propagation}, authors analyze Bitcoin from a networking perspective and show that the propagation delay in the network is the primary cause for blockchain forks. Literature \cite{neudecker2019short} provides an empirical analysis of the announcement and propagation of Bitcoin blocks that caused blockchain forks. Authors in \cite{pass2017analysis} analyze the blockchain protocol in asynchronous networks. In \cite{shahsavari2019theoretical}, authors model the Bitcoin consensus and network protocols to develop the theoretical analysis for fork in Bitcoin network. In \cite{liu2019reducing}, authors try to reduce forks in the blockchain via probabilistic verification.
Nevertheless, existing research efforts mainly focus on the impacting factors from the blockchain system point of view such as block size and network propagation delay.
Along a different line from the previous work, we focus on the computing power perspective and propose a detailed model of temporary fork with heterogeneous computing power.

\textbf{Game Theory in Blockchain Networks.} Blockchain is a distributed, decentralized, public ledger. Due to the decentralization of the blockchain network, game theory \cite{webb2007game} is an ideal modeling tool to analyze the interactions within the blockchain network \cite{liu2019survey}, where evolutionary games \cite{weibull1997evolutionary} can be used to analyzing the dynamic interactions and the evolution of the blockchain networks. The evolutionary game for consensus provision in blockchain networks with shards is investigated in \cite{ni2019evolutionary}. Literature \cite{kim2019mining} proposes an evolutionary game theoretical analysis on block withholding attack in PoW blockchain. 

A closely related work \cite{evolution_pool} develops an evolutionary game to study the dynamic process of mining pool selection. It is worth noting that our work substantially differs from and complements to \cite{evolution_pool} in the following aspects: 1) We propose a detailed mathematical model to characterize the impact of computing power competition of the mining pools on the temporary fork; 2) We investigate the impact of uncle block reward; 3) The mining game in \cite{evolution_pool} is ``fair'', while the mining game in our model can be ``unfair'' under the impact of temporary fork. Thus we obtain different results from \cite{evolution_pool};
4) We investigate the ESS and NSS of the mining game and provide theoretical analysis for a more challenging but general scenario with $M$ mining pools; 5) We obtain several key insights of mining reward and centralization through realistic data based evaluations.

\textbf{Impact of Uncle Block Rewards.} In Ethereum, stale blocks do not have to be discarded but can be referenced as uncle blocks yielding a partial reward called uncle block reward. Several papers have discussed the impact of uncle block reward. In \cite{ritz2018impact}, authors investigate the impact of uncle rewards on selfish mining in Ethereum. In \cite{lerner2016uncle}, the author proposes the uncle mining problem, which is an Ethereum consensus protocol flaw. Authors in \cite{chang2019uncle} evaluate the impact of uncle block reward in block withholding attack. In \cite{liu2020evaluation}, an evaluation of uncle block mechanism effect on Ethereum selfish and stubborn mining combined with an eclipse attack is presented. However, there is few works about the impact of uncle block reward on the evolution of mining pools and the degree of decentralization. Different from previous work, we investigate the impact of uncle block reward from another perspective.

\textbf{Decentralization in Blockchain Networks.} Decentralization is an important feature of the permissionless blockchain network.
Many blockchain-based applications greatly depend on dencetralization \cite{de2016interplay,conoscenti2017peer,kumar2020demistifying}.
There are a few papers studying the decentralization of the blockchain networks. In \cite{cong2019decentralized}, authors study the centralization and decentralization forces in the creation and competition of mining pools in game theoretical approach. Literature \cite{liu2019decentralization} shows that decentralization is vulnerable under the gap game. In \cite{kwon2019impossibility}, authors show that it is impossible to achieve full decentralization in permissionless blockchain networks. In \cite{alzahrani2018towards}, a novel blockchain consensus protocol is proposed based on game theory and randomness to achieve true decentralization. Literature \cite{wu2019information} presents an information entropy method to quantify the degrees of decentralization for blockchain systems.
Literature \cite{gencer2018decentralization} explores the decentralization in Bitcoin and Ethereum networks based on various decentralization metrics and shows that Ethereum is more centralized than Bitcoin, which is indeed consistent with our theoretical analysis. Distinct from the previous work, we study the decentralization of the blockchain networks through the evolutionary game framework and investigate the degree of decentralization under different system settings.

Generally speaking, previous works neglect the disproportionate mining reward of different mining pools caused by the temporary fork phenomenon in the blockchain. Along a different line from the previous researches, we propose a detailed model of the temporary fork phenomenon in the blockchain network and further investigate the evolution of mining pools and the degree of decentralization based on the evolutionary game theory framework.

\section{Conclusion}\label{conclusion}


In this paper, we propose a detailed model of the temporary fork in the blockchain. Combining the mining reward analysis based on the temporary fork model, we find that the mining rewards of the mining pools are disproportionate to their hash rates. Also, the effect of uncle block reward is investigated. Moreover, the evolutionary game of the mining pool evolution under temporary fork is studied.  We theoretically characterize the set of convergent stable states of the evolutionary game, which reveal the long-term trends on the degree of centralization of computing power in the blockchain.  The effectiveness of our models for providing useful insights is corroborated by both numerical simulations and realistic blockchain data analysis. 

For the future study, we are going to integrate other realistic yet challenging factors such as selfish mining and double spending attacks into our model analysis.

\ifCLASSOPTIONcaptionsoff
  \newpage
\fi

\def\authorbio{1}

\ifx\authorbio\undefine

\begin{IEEEbiography}[{\includegraphics[width=1in,height=1.25in,clip,keepaspectratio]{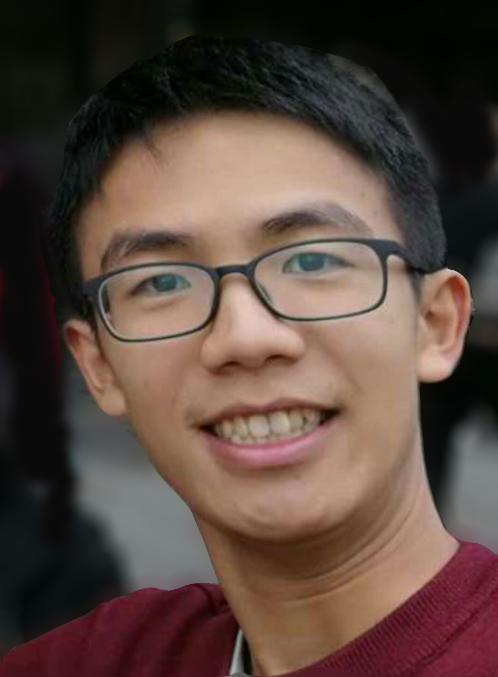}}]{Canhui Chen} 
	is currently pursuing the B.E. degree in software engineering at the School of Data and Computer Science, Sun Yat-sen University, Guangzhou, China. His current research interets include game theory and blockchain system.
\end{IEEEbiography}

\begin{IEEEbiography}[{\includegraphics[width=1in,height=1.25in,clip,keepaspectratio]{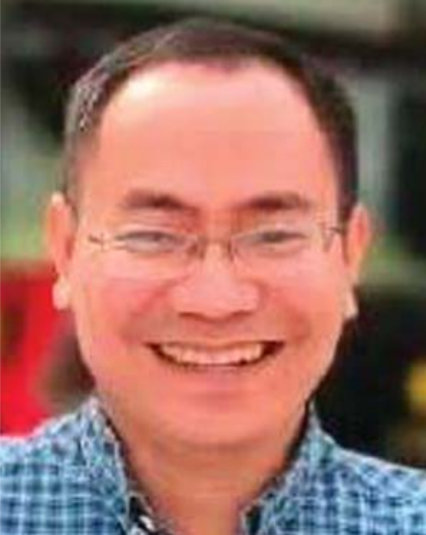}}]{Xu Chen}
	received the Ph.D. degree in information engineering from The Chinese University of Hong Kong in 2012. He has worked as a Post-Doctoral Research Associate at Arizona State University, Tempe, AZ, USA, from 2012 to 2014, and a Humboldt Scholar Fellow at the Institute of Computer Science, University of Goöttingen, Germany, from 2014 to 2016. He is currently a Full Professor with Sun Yat-sen University, Guangzhou, China, and the Vice Director of the National and Local Joint Engineering Laboratory of Digital Home Interactive Applications. He received the prestigious Humboldt Research Fellowship awarded by the Alexander von Humboldt Foundation of Germany, the 2014 Hong Kong Young Scientist Runner-Up Award, the 2017 IEEE Communication Society Asia-Pacific Outstanding Young Researcher Award, the 2017 IEEE ComSoc Young Professional Best Paper Award, the Honorable Mention Award of the 2010 IEEE International Conference on Intelligence and Security Informatics (ISI), the Best Paper Runner-Up Award of the 2014 IEEE International Conference on Computer Communications (INFOCOM), and the Best Paper Award of the 2017 IEEE International Conference on Communications (ICC). He is an Area Editor of the IEEE Open Journal of the Communications Society and an Associate Editor of the IEEE Transactions on Wireless Communications, the IEEE Internet of Things Journal, and the IEEE Journal on Selected Areas in Communications (JSAC) Series on Network Softwarization and Enablers.
\end{IEEEbiography}

\begin{IEEEbiography}[{\includegraphics[width=1in,height=1.25in,clip,keepaspectratio]{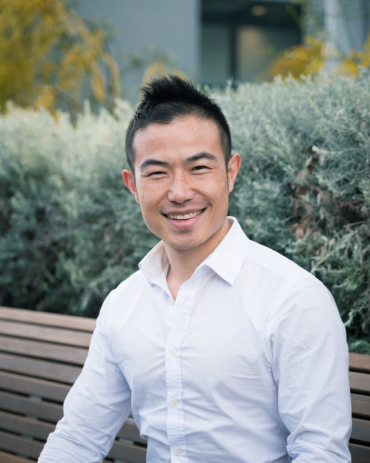}}]{Jiangshan Yu} received the PhD degree from the University of Birmingham, United Kingdom, in 2016. He is currently associate director (research) at Monash Blockchain Technology Centre, Monash University, Australia. Previously, he was a research associate at SnT, University of Luxembourg, Luxembourg. The focus of his research has been on design and analysis of cryptographic protocols, cryptographic key management, blockchain consensus, and ledger-based applications. In particular, His recent research challenges the soundness of the foundational security models and design principles of existing blockchain systems, where the blockchain ecosystem of hundreds of billions of dollars is based upon. He won numerous prestigious awards, including Dean's Research Impact Award (2019) and the Chinese Government Award for Outstanding Scholar Abroad (1 percent worldwide, 2016)
\end{IEEEbiography}

\begin{IEEEbiography}[{\includegraphics[width=1in,height=1.25in,clip,keepaspectratio]{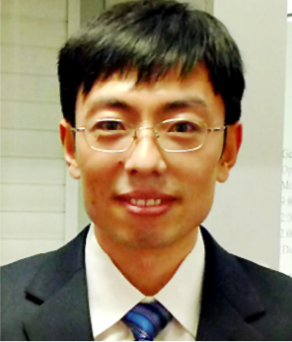}}]{Weigang Wu} received the B.Sc. degree in 1998 and the M.Sc. degree in 2003, both from Xi'an Jiaotong University, China. He received the Ph.D. degree in computer science in 2007 from Hong Kong Polytechnic University. He is currently an associate professor at the department of computer science, Sun Yat-sen University, China.
His research interests include distributed systems and wireless networks, especially cloud computing platforms and ad hoc networks. He has published more than 60 papers in major conferences and journals. He has served as
a member of editorial board of two international journals, Frontiers of Computer Science, and Ad Hoc \& Sensor Wireless Networks. He is also
an organizing/program committee member for many international conferences. He is a member of the IEEE and ACM.
\end{IEEEbiography}

\begin{IEEEbiography}[{\includegraphics[width=1in,height=1.25in,clip,keepaspectratio]{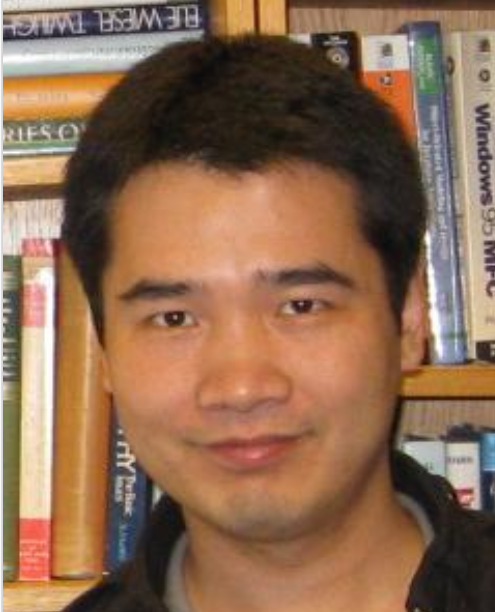}}]{Di Wu} received the B.S. degree
from the University of Science and Technology of
China, Hefei, China, in 2000, the M.S. degree from the Institute of Computing Technology, Chinese Academy of Sciences, Beijing, China, in 2003, and the Ph.D. degree in computer science and engineering from the Chinese University of Hong Kong, Hong Kong, in 2007. He was a Post-Doctoral Researcher with the Department of Computer Science and Engineering, Polytechnic Institute of New York University, Brooklyn, NY, USA, from 2007 to 2009, advised by Prof. K. W. Ross. He is currently a Professor and the
Assistant Dean of the School of Data and Computer Science with Sun Yat-sen University, Guangzhou, China. His research interests include cloud computing, multimedia communication, Internet measurement, and network security.
He was a co-recipient of the IEEE INFOCOM 2009 Best Paper Award. He has served as an Editor of the Journal of Telecommunication Systems (Springer), the Journal of Communications and Networks, Peer-to-Peer Networking and
Applications (Springer), Security and Communication Networks (Wiley), and
the KSII Transactions on Internet and Information Systems, and a Guest Editor of the IEEE TRANSACTIONS ON CIRCUITS AND SYSTEMS FOR VIDEO
TECHNOLOGY. He has also served as the MSIG Chair of the Multimedia Communications Technical Committee in the IEEE Communications Society
from 2014 to 2016. He served as the TPC Co-Chair of the IEEE Global Communications Conference—Cloud Computing Systems, and Networks, and Applications in 2014, the Chair of the CCF Young Computer Scientists and Engineers Forum, Guangzhou from 2014 to 2015, and a member of the
Council of China Computer Federation.
\end{IEEEbiography}


%

\clearpage
\newpage

\else
\fi

\appendices

\allowdisplaybreaks[1]

\section{Analysis of Branches in Temporary Fork}

In this section, we look insight into the branches in temporary fork and show that the case with two competing  branches is a good approximation.
After considering the mining process as a Poisson process with the average block production rate $\lambda$ in section 2.1, based on the property of Poisson process, the probability of generating new blocks during the block confirmation period can be formulated as 
\begin{equation}\nonumber
P^{\Delta}_{n \geq 1} = 1 - e^{-\lambda \tau(s)}.
\end{equation}
And the probability of generating exactly one block during the block confirmation period is
\begin{equation}\nonumber
P^{\Delta}_{n = 1} = \lambda \tau(s) e^{-\lambda \tau(s)}.
\end{equation}
Therefore, under the condition of generating new blocks during the block confirmation period, the probability of generating exactly one block can be calculated according to the property of conditional probability as follows:
\begin{equation}\nonumber
P(\lambda \tau(s)) = \frac{P^{\Delta}_{n = 1}}{P^{\Delta}_{n \geq 1}} = \frac{\lambda \tau(s) e^{-\lambda \tau(s)}}{1 - e^{-\lambda \tau(s)}}.
\end{equation}
As mentioned above in section 2.2, we usually have $\tau(s) < T$ in the realistic blockchain networks and it implies $\lambda \tau(s) < 1$. Indeed, the probability function $P(\lambda \tau(s))$ is monotonically decreasing as $\lambda \tau(s)$ increases. When $\lambda \tau(s) = 0$, $P(\lambda \tau(s))=1$, and even $\lambda \tau(s) = 1$, $P(\lambda \tau(s))> 0.5$. Therefore, when temporary fork happens, the chain is most likely to be forked into two competing branches, which shows that the case with two branches is a good approximation to provide useful insights.  

\begin{figure}[!h]
	\centering
	\subfigure[Frequency of temporary fork in Ethereum]{
		\includegraphics[width=1.5in,height=1.5in]{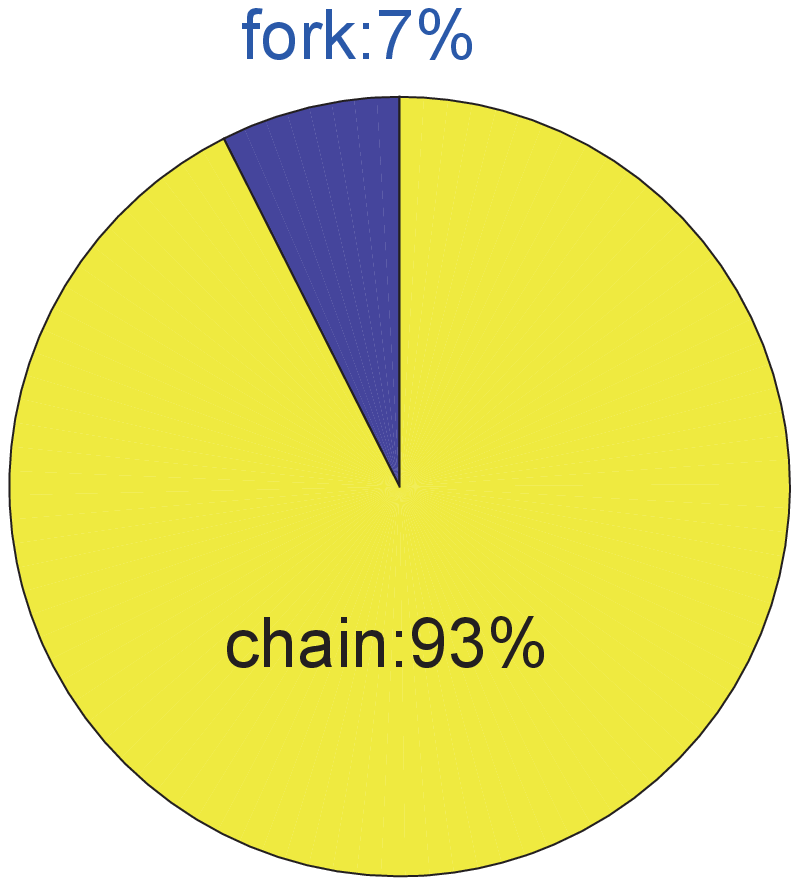}}
	\subfigure[Number of branches in temporary fork ]{\includegraphics[width=1.6in,height=1.5in]{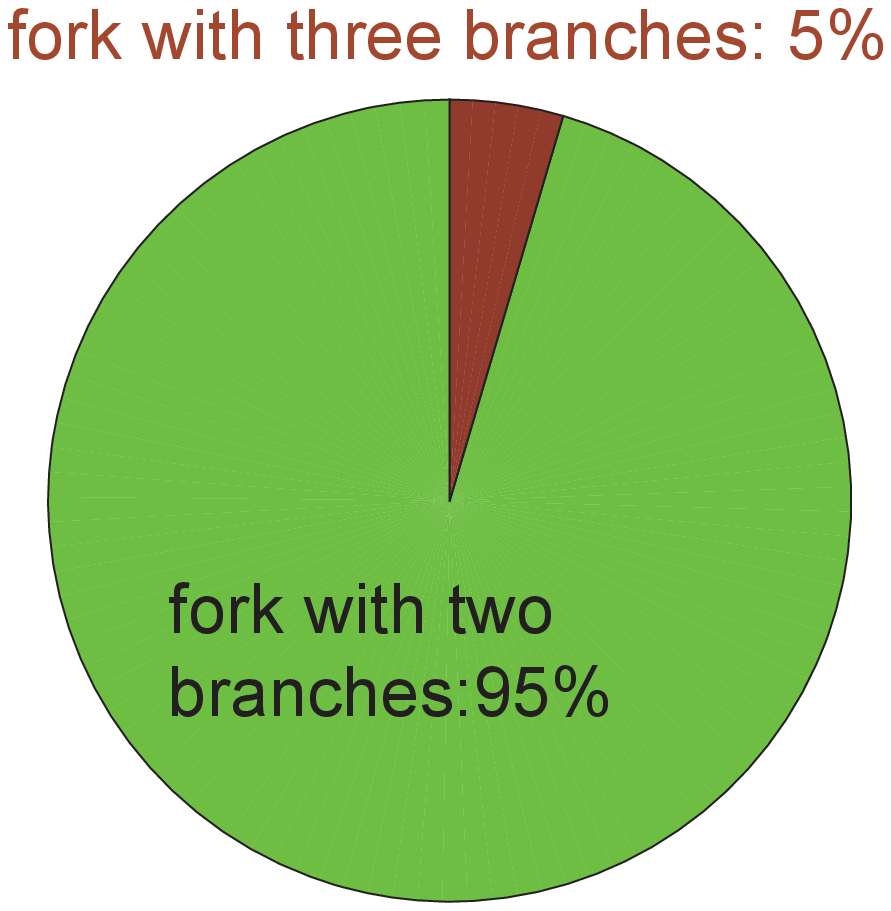}}
	\caption{Illustration of Temporary Fork}
	\label{branches_fig}
\end{figure}
Here, we further investigate the temporary fork phenomenon in realistic blockchain systems. In Bitcoin, the frequency of temporary fork is low, and the temporary fork will always lead to two competing branches \cite{decentralization}. Besides, we collect and analyze data for the first 1,920,000 blocks of the Ethereum blockchain. As shown in Figure \ref{branches_fig}, we find that the ratio of temporary fork in the first 1,920,000 blocks of the Ethereum blockchain is 7.4\%, and that most of the temporary forks lead to two competing branches, which accounts for 95\% of the total, and the temporary forks with three competing branches are rare, which only accounts for 5\%. All in all, the practical insight in realistic blockchain systems is consistent with the theoretical analysis above, which further illustrates the good approximation of considering the case with two branches.

\section{Proof of Theorem \ref{pool_reward}}\label{proof_pool_reward}
The expected mining reward of the mining pool $i$ can be expressed as 
\begin{align}
Y_i &= P_i^{\text{mine a block}} \left( P_i^{\text{uncle}} * \theta R + (1 - P_i^{\text{uncle}}) * R \right) \notag  \\
&= x_i R (1 - (1 - \theta) P_i^{\text{uncle}}),\notag 
\end{align}
where the probability of mining a block is proportional to its hash rate (i.e., $P_i^{\text{mine a block}} = x_i$), and the probability for the mining pool $i$ to mine a uncle block, $P_i^{\text{uncle}}$ is formulated in (\ref{uncle_block}).

\section{Proof of Theorem \ref{same_w}}\label{proof_same_w}
Since $\omega_{1} = \cdots = \omega_{M} = \omega$, we obtain that $x_i = r_i$. Therefore, a miner's expected payoff in pool $i$ in (\ref{game_reward}) can be simplified as 
\begin{small}
\begin{equation}
\begin{aligned}
&y_i(\bm{r}) = \frac{R}{N}\left(1 - (1 - \theta) \frac{1 - e^{-\lambda \tau(s)}}{2} (1 - r_i)\right) - p \omega \notag \\
=& \frac{R}{N} \frac{1 - e^{-\lambda \tau(s)}}{2} (1 - \theta) r_i + \frac{R}{N} - (1 - \theta) \frac{1 - e^{-\lambda \tau(s)}}{2}\frac{R}{N} - p \omega \notag \\
=& a r_i + b \notag,
\end{aligned}
\end{equation}
\end{small}
where $a = \frac{R}{N} \frac{1 - e^{-\lambda \tau(s)}}{2} (1 - \theta)$ and $b = \frac{R}{N} - (1 - \theta) \frac{1 - e^{-\lambda \tau(s)}}{2}\frac{R}{N} - p \omega$.

It is clear that $\mathbf{e}_i, i \in \mathcal{M} = \{1, \cdots, M\}$ are the rest points of ODEs, where $\mathbf{e}_i$ is a vector whose $i$-th component is 1, and the remaining components are 0, i.e., $\mathbf{e}_i = [0 \cdots 1 \cdots 0]^T$, and here we are going to show that they are the ESSs of the game.

Without loss of generality, we will prove that $\bm{e}_1$ is an ESS of the game.
To prove $\textbf{e}_i, i \in \mathcal{M}$ are the ESSs of the game, we need to show that $\textbf{e}_i, \forall i \in \mathcal{M}$ satisfies the condition (4). Without loss of generality, we will prove that $\textbf{e}_1$ is an ESS of the game. Let $\textbf{r}^* = \textbf{e}_1$. Suppose that there exists another population state $\textbf{r}^\prime$ trying to invade the state $\textbf{r}^*$ by attracting a small share $\epsilon \in (0, \bar{\epsilon})$ in the population of miners to switch to $\textbf{r}^{\prime}$. According to condition (4), we need to prove that $\exists \bar{\epsilon} \in (0, 1)$, s.t., $\forall \epsilon \in (0, \bar{\epsilon}), \forall \textbf{r}^{\prime} \in \Delta$ we have
$$
\sum_{i=1}^{M} r_{i}^{*} y_{i}\left((1-\epsilon) \mathbf{r}^{*}+\epsilon \mathbf{r}^{\prime}\right) \geq \sum_{i=1}^{M} r_{i}^{\prime} y_{i}\left((1-\epsilon) \mathbf{r}^{*}+\epsilon \mathbf{r}^{\prime}\right) .
$$
Let $\mathbf{r}^*= \bm{e}_1$. Suppose that there exists another population state $\mathbf{r}'$ trying to invade state $\mathbf{r}^*$ by attracting a small share $\epsilon \in (0, \bar{\epsilon})$ in the population of miners to switch to $\mathbf{r}'$.
	Since $\textbf{r}^* = \textbf{e}_1$, i.e., $r_1^* = 1$ and $r_j^* = 0, j = 2, 3, \ldots, M$, we have 
\begin{equation}\nonumber
\begin{aligned}
&\sum_{i=1}^{M} r_{i}^{*} y_{i}\left((1-\epsilon) \mathbf{r}^{*}+\epsilon \mathbf{r}^{\prime}\right) \\
=& r^*_1 y_1 \left( (1 - \epsilon) \begin{bmatrix}
1 \\ 0 \\ \vdots \\ 0 \\
\end{bmatrix}  + \epsilon \begin{bmatrix}
r'_1 \\ r'_2 \\ \vdots \\ r'_M \\
\end{bmatrix}  \right) \\
=& a( (1 - \epsilon) + \epsilon r'_1 ) + b.
\end{aligned}
\end{equation}
Similarly, we have
\begin{equation}\nonumber
\begin{aligned}
&\sum_{i \in \mathcal{M}} r_{i}^{\prime} y_{i}\left((1-\epsilon) \mathbf{r}^*+\epsilon \mathbf{r}^{\prime}\right) \\
=& \sum_{i \in \mathcal{M}} r_{i}^{\prime} y_i \left( (1 - \epsilon) \begin{bmatrix}
1 \\ 0 \\ \vdots \\ 0 \\
\end{bmatrix}  + \epsilon \begin{bmatrix}
r'_1 \\ r'_2 \\ \vdots \\ r'_M \\
\end{bmatrix}  \right) \\
=& r'_1 \left( a(1 - \epsilon + \epsilon r'_1) + b \right) + \sum_{i=2}^{M} r'_i (a\epsilon r'_i + b) \\
=& a r_1' - a \epsilon r_1'  + a \epsilon \sum_{i=1}^{M}{{r_i'}^2} + b \sum_{i=1}^{M} r_i
\end{aligned}
\end{equation}
Take $\bar{\epsilon} = 1/2$, $\forall \epsilon \in (0, \bar{\epsilon})$, we have that
\begin{align}
&\sum_{i \in \mathcal{M}} r^*_{i} y_{i}\left((1-\epsilon) \mathbf{r}^*+\epsilon \mathbf{r}^{\prime}\right) - \sum_{i \in \mathcal{M}} r_{i}^{\prime} y_{i}\left((1-\epsilon) \mathbf{r}^*+\epsilon \mathbf{r}^{\prime}\right) \notag \\
=& \left( a - a\epsilon(1 - r_1') + b \right) - \left( a r_1' - a \epsilon r_1' + b + a \epsilon \sum_{i=1}^{M}{{r_i'}^2} \right) \notag \\
=& a(1 - r_i') - \epsilon \left( a - 2a r_1' + a \sum_{i=1}^{M}{{r_i'}^2} \right) \notag \\
>& a(1 - r_1') - 2 a \epsilon (1 - r_1') \notag \\
>& 0 . \notag
\end{align} 
Therefore, $\mathbf{e}_1$ is an ESS of the game. Similarly, $\mathbf{e}_i, i \in \{1, \cdots, M\}$ are ESSs of the game. Moreover, we will prove that $\mathbf{e}_i, i \in \{1, \cdots, M\}$ are the only possible ESSs of the system.

Suppose that $\mathbf{r} = [r_1 \cdots r_M]^T$, with at least two non-zero components, i.e., $i \neq j$ and $, r_i \geq r_j > 0$. We will show that $\mathbf{r}$ is not an ESS of the system.

To show that $\mathbf{r}$ is not an ESS of the system, we need to prove $\forall \bar{\epsilon} \in (0, 1)$, $\forall \epsilon \in (0, \bar{\epsilon})$, $\exists \textbf{r}' \in \Delta$, s.t., we have
$$
\sum_{i=1}^{M} r_{i}^{*} y_{i}\left((1-\epsilon) \mathbf{r}^{*}+\epsilon \mathbf{r}^{\prime}\right) < \sum_{i=1}^{M} r_{i}^{\prime} y_{i}\left((1-\epsilon) \mathbf{r}^{*}+\epsilon \mathbf{r}^{\prime}\right) .
$$
Let $\mathbf{r}' = [r_1' \cdots r_M']$, where $r_k' = r_k, \forall k \neq i, k \neq j$, and $r_i' = r_i + \xi$, $r_j' = r_j - \xi$, and $\xi \in (0, \min\{r_i, r_j\})$. Therefore, we have
\begin{align}
&\sum_{i \in \mathcal{M}} r_{i} y_{i}\left((1-\epsilon) \mathbf{r}+\epsilon \mathbf{r}^{\prime}\right) - \sum_{i \in \mathcal{M}} r_{i}^{\prime} y_{i}\left((1-\epsilon) \mathbf{r}+\epsilon \mathbf{r}^{\prime}\right) \notag \\
=& \sum_{i \in \mathcal{M}} \left( r_i - r'_i \right) y_{i}\left((1-\epsilon) \mathbf{r}+\epsilon \mathbf{r}^{\prime}\right) \notag \\
=& \sum_{i \in \mathcal{M}} \left( r_i - r'_i \right) y_{i}\left((1 - \epsilon) \begin{bmatrix}
r_1 \\ \vdots \\ r_i \\ \vdots \\ r_j \\ \vdots \\ r_M \\
\end{bmatrix}  + \epsilon \begin{bmatrix}
r_1 \\ \vdots \\ r_i + \xi \\ \vdots \\ r_j - \xi \\ \vdots \\ r_M \\
\end{bmatrix}\right) \notag \\
=& \xi \left(y_{j}\left((1-\epsilon) \mathbf{r}+\epsilon \mathbf{r}^{\prime}\right) - y_{i}\left((1-\epsilon) \mathbf{r}+\epsilon \mathbf{r}^{\prime}\right)\right) \notag \\
=& \xi \left( (a(r_j - \epsilon \xi) + b) - (a(r_i + \epsilon \xi) + b) \right) \notag \\
=& \xi \left( a (r_j - r_i) - 2a \epsilon \xi  \right) < 0 , \notag
\end{align} 
which contradicts to (4), thus it is not an ESS of the system. Then, the proof of Theorem 2 is completed. 

\section{Proof of Theorem \ref{two_pool_unequal}}\label{proof_two_pool_unequal}
A miner's expected payoff in pool $i \in \{1, 2\}$ can be expressed as 
\begin{small}
	\begin{equation}\nonumber
	\left\{
	\begin{array}{ll}
	\!\!\! y_1(\bm{r}) \!=\! \frac{1}{N r_1} \alpha R \left( 1 - (1 - e^{-\lambda \tau(s)}) (1 - \alpha)^2 (1 - \theta) \right) \!\!- p \omega_{1} , \\
	\!\!\! y_2(\bm{r}) \!=\! \frac{1}{N r_2} (1 - \alpha) R \left( 1 - (1 - e^{-\lambda \tau(s)}) \alpha^2 (1 - \theta) \right) \!\!- p \omega_{2} , \\
	\end{array}
	\right.
	\end{equation}
\end{small}
where $\alpha = \omega_{1} r_1 / (\omega_{1} r_1 + \omega_{2} r_2)$.

Using the constraint $r_1 + r_2 = 1$, we introduce a reduced state $r$ and reduce the ODEs in (\ref{ode}) to a single equation.
\begin{equation}\label{two_pool_ode}
\dot{r}(t) = f(r(t)) = r(t) \left( y(r(t)) - \bar{y}(r(t))\right),
\end{equation}
where
\begin{eqnarray}\nonumber
\left\{
\begin{array}{ll}
r_1 = r, \\
r_2 = 1-r, \\
y(r(t)) = y_1(\bm{r}), \\
\bar{y}(r(t)) = r_1(t) y_1(r) + r_2(t) y_2(r).
\end{array}
\right.
\end{eqnarray}

To solve the ODE (\ref{two_pool_ode}), we need to solve either $r(t) = 0$ or $y_1(\bm{r}(t), \bm{\omega}) - y_2(\bm{r}(t), \bm{\omega}) = 0$.

Indeed, We can easily obtain two rest points of the ODEs, $\{0, 1\}$. 

For the above two-strategy pairwise-contest game, $r^*$ is an ESS if and only if it is an asymptotically stable fixed point in the replicator dynamics of the equation (\ref{two_pool_ode}). Since the replicator dynamics (\ref{two_pool_ode}) is continuous-time, $r^*$ is asymptotically stable fixed point if $\frac{\partial \dot{r}(r^*)}{\partial r} < 0$.
When $r^* = 1$, we obtain that 
\begin{small}
\begin{equation}\nonumber
\frac{\partial \dot{r}}{\partial r}(r^*) \!= p \omega_{1} - p \omega_{2} - \frac{R}{N} + \frac{R \omega_{2}}{N \omega_{1}} + \frac{R \left( 1 - e^{-\lambda \tau(s)} \right) \omega_{2} (\theta - 1) }{N \omega_{1}}.
\end{equation}
\end{small}
Thus $r^* = 1$ is an ESS when 
\begin{equation}\nonumber
p N \omega_{1} < R + \frac{R \left( 1 - e^{-\lambda \tau(s)} \right) \omega_{2} (1 - \theta)}{\omega_{1} - \omega_{2}}.
\end{equation}
When $r^* = 0$, we obtain that 
\begin{small}
\begin{equation}\nonumber
\frac{\partial \dot{r}}{\partial r}(r^*) = p \omega_{2} - p \omega_{1} - \frac{R}{N} + \frac{R \omega_{1}}{N \omega_{2}} + \frac{R \left( 1 - e^{-\lambda \tau(s)} \right) \omega_{2} (\theta - 1) }{N \omega_{2}}.
\end{equation}
\end{small}
Thus $r^* = 0$ is an ESS when 
\begin{equation}\nonumber
p N \omega_{2} > R - \frac{R \left( 1 - e^{-\lambda \tau(s)} \right) \omega_{1} (1 - \theta)}{\omega_{1} - \omega_{2}}.
\end{equation}

Other possible fixed points of the ODEs can be obtained by solving the equation $y_1(r) - y_2(r) = 0$, which is equivalent to solving the following cubic equation. 
\begin{equation}\label{cubic}
a r^3 + b r^2 + c r + d = 0 ,
\end{equation}
where 
\begin{small}
\begin{equation}\nonumber
\left\{
\begin{array}{ll}
a = N p (\omega_{1} - \omega_{2})^4, \\
b = R (\omega_{1} - \omega_{2})^3 + R (1 - e^{-\lambda \tau(s)}) (1-\theta) \omega_{1} \omega_{2} (\omega_{1} - \omega_{2})\\
\qquad - 3Np\omega_{2}(\omega_{1} - \omega_{2})^3 , \\
c = 2R\omega_{2}(\omega_{1} - \omega_{2})^2 + 2 R (1 - e^{-\lambda \tau(s)}) (1-\theta) \omega_{1} \omega_{2}^2 \\
\qquad- 3N p \omega_{2}^2 (\omega_{1} - \omega_{2})^2, \\
d = -\omega_{2}^2 (R (1 \!-\! e^{-\lambda \tau(s)}) (1 \!-\! \theta) \omega_{1} \!-\! (R \!-\! N p \omega_{2}) (\omega_{1}\! -\! \omega_{2}).
\end{array}
\right.
\end{equation}
\end{small}

Indeed, $\exists r^* \in [0, 1]$, s.t., $y_1(r^*) = y_2(r^*)$ and $r^*$ is an asymptotically stable fixed point of the replicator dynamics under the following conditions
\begin{eqnarray}\nonumber
\left\{
\begin{array}{ll}
p N \omega_{1} \geq R + \frac{R \left( 1 - e^{-\lambda \tau(s)} \right) \omega_{2} (1 - \theta)}{\omega_{1} - \omega_{2}},\\
p N \omega_{2} \leq R - \frac{R \left( 1 - e^{-\lambda \tau(s)} \right) \omega_{1} (1 - \theta)}{\omega_{1} - \omega_{2}},
\end{array}
\right.
\end{eqnarray}

Denote $h(r) = y_1(r) - y_2(r)$, then we have
\begin{small}
\begin{equation}\nonumber
\left\{
\begin{array}{ll}
\!\!\! h(0) = p \omega_{1} - p \omega_{2} + \frac{R}{N} - \frac{R \omega_{1}}{N \omega_{2}} - \frac{R \left( 1 - e^{-\lambda \tau(s)} \right) \omega_{2} (\theta - 1) }{N \omega_{2}} \geq 0, \\
\!\!\! h(1) = p \omega_{1} - p \omega_{2} - \frac{R}{N} + \frac{R \omega_{2}}{N \omega_{1}} + \frac{R \left( 1 - e^{-\lambda \tau(s)} \right) \omega_{2} (\theta - 1) }{N \omega_{1}} \leq 0.
\end{array}
\right.
\end{equation}
\end{small}

Then, according to the intermediate value theorem, we have $\exists r^* \in [0, 1]$, s.t. $h(r^*) = 0$ and $h^{(k)}(r^*) \leq 0, \forall k \in \{1, 2, \cdots\}$, which $h^{(k)}$ denotes the $k$-th derivative of $h(r)$.
Moreover, we have
\begin{eqnarray}\nonumber
\begin{aligned}
\frac{\partial \dot{r}}{\partial r}(r^*) &= y_1 - \bar{y} + r_1 \left( \frac{\partial y_1}{\partial r_1} - \frac{\partial \bar{y}}{\partial r_1} \right) \\
&= (1 - r^*) \left( \frac{\partial y_1}{\partial r}(r^*) - \frac{\partial y_2}{\partial r}(r^*) \right) \\
&= (1 - r^*_1) h'(r^*) \leq 0
\end{aligned} 
\end{eqnarray}

Above all, $r^*$ is an asymptotically stable fixed point of the replicator dynamics, thus an ESS of the game.

\section{Proof of Theorem \ref{uncle_reward_limit_two}}\label{proof_uncle_reward_limit_two}
When the uncle block reward is large enough, i.e., $\theta \rightarrow 1$, according to the condition in (\ref{r_equal_1}) (\ref{r_equal_0}), we obtain that the ESS of the game is $r^* = 1$, if $R \geq p N \omega_{1}$ and that the ESS of the game is $r^* = 0$, if $R \leq p N \omega_{2}$

When $\theta \rightarrow 1$, the solution of the cubic equation in (\ref{cubic}) is $r^* = \frac{R - \omega_{2} p N}{p N (\omega_{1} - \omega_{2})}$ and we have that 
\begin{equation}\nonumber
\frac{\partial \dot{r}}{r}(r^*) = \frac{(R - N p \omega_{1})(R - N p \omega_{2})}{NR} + (1-\theta) h(r^*)
\end{equation}
where 
\begin{small}
\begin{equation}\nonumber
\begin{aligned}
h(r^*) \!=\! \frac{(1 \!-\! e^{-\lambda \tau(s)}) N^2p^3 \omega_{1}^2 \omega_{2}^2}{R^3(\omega_{1} \!-\! \omega_{2})^2}  (2R\omega_{1} \!+\! 2R \omega_{2} \!-\! 3Np \omega_{1} \omega_{2} \!+\! R^4)
\end{aligned}
\end{equation}
\end{small}
Therefore, when $\theta \rightarrow 1$, we have $\frac{\partial \dot{r}}{r}(r^*) < 0$. $r^*$ is asymptotically stable point of the ode, thus is an ESS of the system.

\section{Proof of Theorem \ref{general_case_limit}}\label{proof_general_case_limit}

The miner expected reward in mining pool $i$ in (\ref{game_reward}) can be rewritten as follows.
\begin{equation}\nonumber
y_i = \frac{R \omega_{i}}{N \sum_{k=1}^{M}{\omega_{k} r_k}} - p \omega_{i} + (1- \theta) \frac{1 - e^{-\lambda \tau(s)}}{2} g(\bm{r}, \bm{\omega})
\end{equation}
where
\begin{eqnarray}\nonumber
\begin{aligned}
&g(\bm{r}) = -\frac{R x_i}{N r_i} \left( \sum_{j \neq i}{x_j (1 - x_i + x_j)} \right)
\end{aligned}
\end{eqnarray}

Therefore, if the network propagation time is negligible, i.e., $\tau(s) = 0$, or the uncle reward is large enough, i.e., $\theta = 1$, the miner expected reward in mining pool $i$ can be expressed as 

\begin{equation}\nonumber
y_i = \frac{R \omega_{i}}{N \sum_{k=1}^{M}{\omega_{k} r_k}} - p \omega_{i}
\end{equation}

\subsection{Proof for the Case $R \geq p N \omega_{1}$}

Let $\mathbf{r}^*=[r_1^* \ \cdots \ r_M^* ]$, where $r_1^* = 1$ and $r_j^* = 0, \forall j \in \{2, \cdots, M\}$. Suppose that there exists another population state $\mathbf{r}'$ trying to invade state $\mathbf{r}^*$ by attracting a small share $\epsilon \in (0, \bar{\epsilon})$ in the population of miners to switch to $\mathbf{r}'$, then we have
\begin{align}\nonumber
&\sum_{i=1}^{M} r_{i}^* y_{i}\left((1-\epsilon) \mathbf{r}^*+\epsilon \mathbf{r}^{\prime}\right) -\sum_{i=1}^{M} r_{i}^{\prime} y_{i}\left((1-\epsilon) \mathbf{r}^*+\epsilon \mathbf{r}^{\prime}\right) \notag \\
=& \sum_{i=1}^{M} { (r_i^* - r_i') y_{i}\left((1-\epsilon) \mathbf{r}^*+\epsilon \mathbf{r}^{\prime}\right) } \notag \\
=& \frac{R \omega_{1}}{N \left( (1 - \epsilon) \omega_{1} + \epsilon \sum_{k=1}^{M}{\omega_{k} r_k'} \right) } - p \omega_{1} \notag \\
&\quad - \sum_{i=1}^{M}{r_i' \left( \frac{R \omega_{i}}{ N \left( (1-\epsilon) \omega_{1} + \epsilon \sum_{k=1}^{M}{\omega_{i} r_i'} \right) } - p \omega_{i} \right) } \notag \\
=& \frac{R}{N} \frac{\omega_{1} - \sum_{i=1}^{M}{\omega_{i} r_i'} }{(1-\epsilon) \omega_{1} + \epsilon \sum_{k=1}^{M}{\omega_{i} r_i'} } - p \left( \omega_{1} - \sum_{i=1}^{M}{ \omega_{i} r_i' } \right) \notag \\
=& \left( \omega_{1} - \sum_{i=1}^{M}{\omega_{i} r_i'} \right) \left( \frac{R}{N \left( (1 - \epsilon) \omega_{1} + \epsilon \sum_{k=1}^{M}{\omega_{k} r_k'} \right) } - p \right) \notag \\
\geq& \left( \omega_{1} - \sum_{i=1}^{M}{\omega_{i} r_i'} \right) \left( \frac{p \omega_{1}}{ \left( (1 - \epsilon) \omega_{1} + \epsilon \sum_{k=1}^{M}{\omega_{k} r_k'} \right) } - p \right) \notag\\
>&0. \notag
\end{align}
Thus, $\mathbf{r}^*$ is an ESS of the game.

\subsection{Proof for the Case $R \leq p N \omega_{M}$}

Let $\mathbf{r}^*=[r_1^* \ \cdots \ r_M^* ]$, where $r_M^* = 1$ and $r_j^* = 0, \forall j \in \{1, \cdots, M-1\}$. Suppose that there exists another population state $\mathbf{r}'$ trying to invade state $\mathbf{r}^*$ by attracting a small share $\epsilon \in (0, \bar{\epsilon})$ in the population of miners to switch to $\mathbf{r}'$, then we have

\begin{align}\nonumber
&\sum_{i=1}^{M} r_{i}^* y_{i}\left((1-\epsilon) \mathbf{r}^*+\epsilon \mathbf{r}^{\prime}\right) -\sum_{i=1}^{M} r_{i}^{\prime} y_{i}\left((1-\epsilon) \mathbf{r}^*+\epsilon \mathbf{r}^{\prime}\right) \notag \\
=& \sum_{i=1}^{M} { (r_i^* - r_i') y_{i}\left((1-\epsilon) \mathbf{r}+\epsilon \mathbf{r}^{\prime}\right) } \notag \\
=& \frac{R \omega_{M}}{N \left( (1 - \epsilon) \omega_{M} + \epsilon \sum_{k=1}^{M}{\omega_{k} r_k'} \right) } - p \omega_{M} \notag \\
&\quad - \sum_{i=1}^{M}{r_i' \left( \frac{R \omega_{i}}{ N \left( (1-\epsilon) \omega_{M} + \epsilon \sum_{k=1}^{M}{\omega_{i} r_i'} \right) } - p \omega_{i} \right) } \notag \\
=& \frac{R}{N} \frac{\omega_{M} - \sum_{i=1}^{M}{\omega_{i} r_i'} }{(1-\epsilon) \omega_{M} + \epsilon \sum_{k=1}^{M}{\omega_{i} r_i'} } - p \left( \omega_{M} - \sum_{i=1}^{M}{ \omega_{i} r_i' } \right) \notag \\
=& \left( \omega_{M} - \sum_{i=1}^{M}{\omega_{i} r_i'} \right) \!\! \left( \frac{R}{N \left( (1 - \epsilon) \omega_{M} + \epsilon \sum_{k=1}^{M}{\omega_{k} r_k'} \right) } - p \right) \notag \\
\geq& \left( \omega_{M} - \sum_{i=1}^{M}{\omega_{i} r_i'} \right) \!\! \left( \frac{p \omega_{M}}{ \left( (1 - \epsilon) \omega_{M} + \epsilon \sum_{k=1}^{M}{\omega_{k} r_k'} \right) } - p \right) \notag\\
>&0. \notag
\end{align} 
Thus, $\mathbf{r}^*$ is an ESS of the game.

\subsection{Proof for the Case $pN\omega_{M} < R < pN\omega_{1}$}

We firstly prove that $\Delta^{\text{NSS}} = \{ \mathbf{r}^* | \sum_{i=1}^{M}{ r_i^* \omega_{i} = \frac{R}{p N} }, \mathbf{r}^* \in \Delta \}$ is the set of NSSs of the game.

Let $\mathbf{r}^*=[r_1^* \ \cdots \ r_M^* ]$, with the following constraint
\begin{equation}\nonumber
\sum_{k=1}^{M}{\omega_{k} r_k^*}  = \frac{R}{Np}.
\end{equation}

Suppose that there exists another population state $\mathbf{r}'$ trying to invade state $\mathbf{r}^*$ by attracting a small share $\epsilon \in (0, \bar{\epsilon})$ in the population of miners to switch to $\mathbf{r}'$, then we have
\begin{small}
\begin{equation}\nonumber
\begin{aligned}
&\sum_{i=1}^{M} r_{i}^* y_{i}\left((1-\epsilon) \mathbf{r}^*+\epsilon \mathbf{r}^{\prime}\right) -\sum_{i=1}^{M} r_{i}^{\prime} y_{i}\left((1-\epsilon) \mathbf{r}^*+\epsilon \mathbf{r}^{\prime}\right) \\
=& \sum_{i=1}^{M} { (r_i^* - r_i') y_{i}\left((1-\epsilon) \mathbf{r}^*+\epsilon \mathbf{r}^{\prime}\right) } \\
=& \sum_{i=1}^{M} { (r_i^* - r_i') \left( \frac{R \omega_{i}}{(1 - \epsilon) \frac{R}{N p} + \epsilon \sum_{k=1}^{M}{\omega_{k} r_k'} } - p \omega_{i} \right) } \\
=& \sum_{i=1}^{M} { \left( \frac{R}{Np} \!-\! \sum_{k=1}^{M}{\omega_{k} r_k'} \right) \! \left( \frac{R }{(1 - \epsilon) \frac{R}{N p} + \epsilon \sum_{k=1}^{M}{\omega_{k} r_k'} } - p  \right)  } \\
\geq & 0.
\end{aligned} 
\end{equation}
\end{small}
The equality holds if and only if $\sum_{k=1}^{M}{\omega_{k} r_k'} = \frac{R}{Np}$, thus $\mathbf{r}$ is a NSS of the system. Therefore, $\Delta^{\text{NSS}} = \{ \mathbf{r}^* | \sum_{i=1}^{M}{ r_i^* \omega_{i} = \frac{R}{p N} } \}$ is the set of NSSs of the game.

We further study the asymptotically stable fixed points of the replicator dynamics. For the ODEs in (\ref{ode}) in the replicator dynamics, by using the constraint $\sum_{i=1}^{M}{r_i}=1$, we can introduce a reduced state vector $\mathbf{r} = (r_1, r_2, \cdots, r_{M-1})$ and reduce the number of equations in (\ref{ode}) to $M-1$.
\begin{equation}\nonumber
\mathbf{r} = (r_1, r_2, \cdots, r_{M-1}).
\end{equation}
Thus, we can write the dynamical system more compactly in vector format as 
\begin{equation}\nonumber
\dot{\mathbf{r}} = f(\mathbf{r}).
\end{equation}
Then the Jacobian matrix of the system can be obtained as
\begin{equation}\label{jacobian_same}
J(\mathbf{r})=
\begin{bmatrix}
{\frac{\partial f_{1}(\mathbf{r})}{\partial r_{1}}} & \cdots & {\frac{\partial f_{1}(\mathbf{r})}{\partial r_{M-1}}} \\ 
\vdots & \cdots & \vdots \\
{\frac{\partial f_{M-1}(\mathbf{r})}{\partial r_{1}}} & \cdots & {\frac{\partial f_{M-1}(\mathbf{r})}{\partial r_{M-1}}}
\end{bmatrix} ,
\end{equation}
where
\begin{equation}\nonumber
\begin{cases}
{\frac{\partial f_{i}(\mathbf{r})}{\partial r_{i}}} = y_i - \bar{y} + r_i \left( \frac{\partial y_i}{\partial r_i} - \frac{\partial \bar{y}}{\partial r_i} \right), & \forall i \in \{1, \cdots, M\},\\
{\frac{\partial f_{i}(\mathbf{r})}{\partial r_{j}}} = -r_i \frac{\partial \bar{y}}{\partial r_j}, & \forall i \neq j .
\end{cases} 
\end{equation}
And we have that
\begin{equation}\nonumber
\frac{\partial \bar{y}}{\partial r_j} = \sum_{k=1}^{M}{r_k \frac{\partial y_k}{\partial r_j}} + y_j, \quad \forall j \in \{1, \cdots, M\}.
\end{equation}
Moreover, when $\sum_{i=1}^{M}{\omega_{k} r_k} = \frac{R}{Np}$, we obtain that
\begin{equation}\nonumber
y_i = \frac{R \omega_{i}}{N \sum_{k=1}^{M}{\omega_{k} r_k}} - p \omega_{i} = 0, \quad \forall i \in \{1, \cdots, M\}.
\end{equation}
Therefore, the elements of the Jacobian matrix in (\ref{jacobian_same}) are derived as follows.
\begin{footnotesize}
	\begin{equation}\nonumber
	\begin{cases}
	{\frac{\partial f_{i}(\mathbf{r})}{\partial r_{i}}} = r_i p (\omega_{i} - \omega_{M}) \left( -\frac{N p \omega_{i}}{R} + 1 \right), & \forall i \in \{1, \cdots, M-1\},\\
	{\frac{\partial f_{i}(\mathbf{r})}{\partial r_{j}}} = r_i p (\omega_{j} - \omega_{M}), & \forall i \neq j .
	\end{cases} 
	\end{equation}
\end{footnotesize}
Further, the Jacobian matrix can be rewritten as 
\begin{footnotesize}
	\begin{equation}\nonumber
	\begin{aligned}
	J(\mathbf{r})&=
	-\frac{Np^2}{R}
	\begin{bmatrix}
	r_1 \omega_{1} (\omega_{1} - \omega_{M})  &   & \\ 
   &\ddots & \\
	  &   & r_{M-1} \omega_{M-1}(\omega_{M-1} - \omega_{M}) 
	\end{bmatrix} \\
	&\quad + p
	\begin{bmatrix}
	r_1 (\omega_{1} - \omega_{M})  &  \cdots & r_1 (\omega_{M-1} - \omega_{M})\\ 
	r_2 (\omega_{1} - \omega_{M}) & \cdots & r_2 (\omega_{M-1} - \omega_{M}) \\
	$\vdots$   &\ddots & \vdots \\
	r_{M-1} (\omega_{1} - \omega_{M})  &  \cdots & r_{M-1} (\omega_{M-1} - \omega_{M}) 
	\end{bmatrix} \\
	&=pC(A+ a a^T) B,
	\end{aligned}
	\end{equation}
\end{footnotesize}
where
\begin{eqnarray}\nonumber
\begin{aligned}
&C = \text{diag}(r_1, r_2, \cdots, r_{M-1}), \\
&A = -\frac{Np}{R} \text{diag}(\omega_{1}, \omega_{2}, \cdots, \omega_{M-1}), \\
&B = \text{diag}(\omega_{1} - \omega_{M}, \omega_{2} - \omega_{M}, \cdots, \omega_{M-1} - \omega_{M}), \\
&a = [1\ 1\ \cdots\ 1]^T. \\
\end{aligned}
\end{eqnarray}
The leading principal minor of order $k$ of the Jacobian matrix denoted as $D_k$ is obtained as follows.
\begin{equation}\nonumber
D_k = p^k \det{(C_k)} \det{(A_k + a_k a_k^T)} \det{(B_k)},
\end{equation}
Using Cauchy's formula for the determinant of a rank-one perturbation \cite{horn2012matrix}, we obtain that
\begin{equation}\nonumber
\begin{aligned}
&\det{(A_k + a_k a_k^T)} \\
=& \det{A_k} + a_k^T\left(\text{adj}{A_k}\right) a_k \\
=& \det{A_k}\left(1 + a_k^T A_k^{-1} a_k\right) \\
=& \det{A_k}\left(1 - \frac{R}{Np} \sum_{i=1}^{k}\frac{1}{\omega_{i}}\right) \\
=& (-1)^k\left(\frac{Np}{R}\right)^k\left(\prod_{i=1}^{k}{\omega_{k}}\right) \left(1 - \frac{R}{Np} \sum_{i=1}^{k}\frac{1}{\omega_{i}}\right).
\end{aligned}
\end{equation}
Therefore we have
\begin{small}
\begin{equation}\nonumber
D_k =  (-1)^k\left(\! \frac{Np^2}{R} \! \right)^k \!\!  \left(\prod_{i=1}^{k}{\omega_{i}r_i (\omega_i - \omega_{i-1})}\right) \left(1 - \frac{R}{Np} \sum_{i=1}^{k}\frac{1}{\omega_{i}}\right).
\end{equation}
\end{small}
Additionally, we hold the following inequality
\begin{eqnarray}\nonumber
\begin{aligned}
1 - \frac{R}{Np} \sum_{i=1}^{k}\frac{1}{\omega_{i}} &\geq 1 - \frac{R}{Np} \sum_{i=1}^{M}\frac{1}{\omega_{i}} \\
&= 1 - \sum_{j=1}^{M} r_j \omega_{j}\sum_{i=1}^{M}\frac{1}{\omega_{i}} \\
&= 1 - \left( \sum_{i=1}^{M}{r_i} + \sum_{i=1}^{M}\sum_{j\neq i}{\frac{r_j \omega_{j}}{\omega_{i}}} \right) \\
&= \sum_{i=1}^{M}\sum_{j\neq i}{\frac{r_j \omega_{j}}{\omega_{i}}} > 0.
\end{aligned}
\end{eqnarray} 
Therefore, if $r_i > 0, \forall i \in \{1, \cdots, M\}$, the Jacobian matrix's odd principal minors are negative and its even principal minors are positive, thus the Jacobian matrix is negative definite, thus, it is an asymptotically stable fixed point in the replicator dynamics.

\bibliographystyle{IEEEtran}
\bibliography{IEEEabrv,IEEE_TNSE_Mining_ref}

%
%







\end{document}